\documentclass[a4paper,12pt]{article}
\usepackage{epsfig}
\usepackage{subfigure}
\usepackage{amsmath}
\usepackage{epic}
\usepackage{amssymb}
\usepackage{array}
\usepackage{fancybox}
\newcommand{\modu}[1]{(\negthickspace\mod{#1})}
\newcommand{\nnnegthickspace}{\negthickspace\negthickspace\negthickspace}
\begin{document}
\setcounter{equation}{0}
\setcounter{section}{0}
\title{ \bf Infinite chain of $N$ different deltas: A simple model
for a quantum wire}
\author{{\bf Jose M. Cerver\'o\footnote{\textbf{cervero@sonia.usal.es}. Author to whom all
correspondence should be addressed}
\  and Alberto Rodr\'{\i}guez} \\ {\small \bf
F\'{\i}sica Te\'orica. Facultad de Ciencias. Universidad de Salamanca} \\
{\small \bf 37008. Salamanca. Spain}}
\date{}
\maketitle
\begin{abstract}  
\noindent We present the exact diagonalization of the Schr\"odinger
operator corresponding to a
periodic potential with $N$ deltas of different couplings, for arbitrary
$N$. This basic structure can
repeat itself an infinite number of times. Calculations of band
structure can  be performed with a high degree of
accuracy for an infinite chain and of the correspondent eigenlevels in the
case of a random chain. The main physical motivation is to modelate
quantum wire band structure and the calculation of
the associated density of states. These quantities show the fundamental properties 
we expect for periodic structures although for low energy the band gaps follow
unpredictable patterns. In the case of random chains we find 
Anderson localization; we analize also the role of the eigenstates in the localization 
patterns and find clear signals of fractality in the conductance. In spite of the 
simplicity of the model many of the salient features expected in a quantum wire are well 
reproduced.  
\end{abstract}
\vskip 0.4cm
$\qquad\quad${\bf PACS Numbers: 03.65.-w: Quantum Mechanics}

$\qquad\qquad\qquad\qquad\qquad${\bf 71.23.An: Theories and Models; Localized States}

$\qquad\qquad\qquad\qquad\qquad${\bf 73.21.Hb: Quantum Wires}
\vskip 0.3 true in
\newpage

Quantum wires represent the dreamed idea of make a conductor wire as small
as a molecule. The
idea that macroscopic devices we have been using for years can actually be
built in nature at the nanoscale size
goes back to Feynman and is the basis of all work currently carried out in
the field of Quantum Electronics.
This area of research is not only interesting in itself from the
fundamental point of view but has also profound
implications in applied physics and material sciences as hundreds of
experiments are being carried out nowdays with
a high degree of success.  The purpose of this paper is to show that some
of the ideas underlying the actual development
of different types of quantum wires can actually be modelled in quite a
simple manner using elementary quantum mechanics.
This is specially important from our point of view as it provides a bridge
between current fundamental research and basic
concepts of quantum mechanics which are usually the subject of graduate
standard programs.

The aim to modelate a simple one-dimensional solid in order to study its
band structure goes back to Kronig and Penney
in the thirties \cite{KP} but since then much work has been done along
the lines of this first seminal reference. A
sample of the variety of models that can be constructed with the same ideas
can be found in \cite{LM}. As the main
interest of almost all of these authors was band theory, the techniques
used in all these papers were mainly addressed
to semiconductor physics \cite{altman}. Much more recently a revival of the
same models and techniques \cite{SZML} has
arisen as a consequence of the interest in truly theoretical and
experimental one-dimensional physical systems from which
quantum wires are just only one example \cite{CDEK}. Right now the research
in molecular conductances, one-dimensional
metallic rings and other devices of the same sort lends support to the idea
of generalizing old methods yielding exact
analitic solutions coupled to the use of desk-top computer algebra.

Here firstly we shall  solve analitically the band structure of {\bf  an
infinite periodic chain of delta potentials each one with a
different coupling inside the primitive cell} paying mostly attention to the mathematical aspects.
After studing the band structure of the
chain one can introduce randomness boosted by quantum fluctuations in order
to account for localization. Fortunately the
model seems rich enough to yield more information such as the
generalization of the Saxon-Hutner conjecture \cite{SH}
and even scaling exhibited as fractal behaviour of the conductance \cite{Hegger}.
The Paper will be organized as follows. In Section I we shall present the analytic solution of
{\bf the periodic case} of infinite different delta potentials.  The band structure of
this one dimensional Periodic Potential will also be briefly discussed. In Section II we turn our attention to
the case of {\bf random arrays} of one dimensional delta potentials with different couplings. 
Here we discuss the density of states using the functional equation method and classify the different 
types of localization appearing when disorder is present. As we have been able to increase our understanding of
localization to the extent of detecting universality effects, we shall entirely devote the Section III to
the discussion of fractality in the conductance and the different checkings we have managed to perform in order
to ascertain ourselves and try to convince the reader that the effect is present even in this extremely simple
model. We close with a Section of Conclusions.
\section{Periodic Array}
Let us consider an electron in a periodic one dimensional chain of atoms
modelled by the potential constituted by an
array of $N$ delta functions each one with its own coupling $e_{i}^2$,
($i=1,2,...N$). After finishing the $N$-array, the
structure repeats itself an infinite number of times. The number of {\it
species}  $N$, can be arbitrarily large but
finite. The case $N$=1 is an old textbook
exercise but may be convenient to be revisited
\cite{FLUG} for taking a full profit of our general results. The
generalization can thus be followed in a more
straightforward manner. The relevant primitive cell for $N=2$ can be
represented for the following set of wavefunctions:
\begin{align*}
  \text{\sl \small{primitive cell}}\, &\begin{cases} \Psi_1(x) \,=\, A_1e^{ikx} +
                B_1e^{-ikx}  \qquad & 0\leq x <a\\
                \Psi_2(x)\,=\, A_2e^{ik(x-a)} + B_2e^{-ik(x-a)}  \qquad & a \leq x  <2a
                \end{cases}\\
    & \quad \Psi_3(x) \,=\, e^{2iQa} \left[ A_1e^{ik(x-2a)} + B_1e^{-ik(x-2a)}\right] 
\end{align*}
The matrix relating the amplitudes of the above wave functions  for this
$N=2$ case can be written as:
\begin{equation}
    \begin{pmatrix}
    e^{ika} & e^{-ika} & \begin{picture}(0,0)(0,70)\dottedline{4}(0,15)(0,75)\end{picture}& -1 & -1 \\
    -ike^{ika} & ike^{-ika} & & \big(ik-{2\over a_1}\big) & -\big(ik+{2\over a_1}\big) \vspace{-2mm}\\ 
    \begin{picture}(0,0)(100,0)\dottedline{4}(80,0)(335,0)\end{picture} \\
    -\mathcal{F}_2 & -\mathcal{F}_2 & & e^{ika} & e^{-ika} \\
    \big(ik-{2\over a_2}\big)\mathcal{F}_2 & -\big(ik+{2\over
a_2}\big)\mathcal{F}_2 & & -ike^{ika} & ike^{-ika} 
    \end{pmatrix}
    \label{ec:det2}
\end{equation}\\ 
It is trivial to generalize these two steps to the case of three
species (i.e. $N=3$). One can equally write
the correspondent matrix in the form:
\begin{equation}
    \begin{pmatrix}
    e^{ika} & e^{-ika}
&\begin{picture}(0,0)(0,90)\dottedline{4}(0,0)(0,95)\end{picture}  & -1 &
-1 &\begin{picture}(0,0)(0,90)\dottedline{4}(0,0)(0,95)\end{picture} & 0 & 0 \\
    -ike^{ika} & ike^{-ika} & & \big(ik-\frac{2}{a_1}\big) &
    -\big(ik+\frac{2}{a_1}\big) & & 0 & 0 \vspace{-2mm}\\
    \begin{picture}(0,0)(20,0)\dottedline{4}(0,0)(380,0)\end{picture}\\
    0 & 0 & & e^{ika} & e^{-ika} & & -1 & -1 \\
    0 & 0 & & -ike^{ika} & ike^{-ika} & & \big(ik-\frac{2}{a_2}\big) &
    -\big(ik+\frac{2}{a_2}\big) \vspace{-2mm} \\
    \begin{picture}(0,0)(20,0)\dottedline{4}(0,0)(380,0)\end{picture}\\
    -\mathcal{F}_3 & -\mathcal{F}_3 & & 0 & 0 & & e^{ika} & e^{-ika} \\
    \big(ik-\frac{2}{a_3}\big)\mathcal{F}_3 &
-\big(ik+\frac{2}{a_3}\big)\mathcal{F}_3 & & 0 & 0 & & -ike^{ika} & ike^{-ika}
    \end{pmatrix}
    \label{ec:det3}
\end{equation}
where in all of the above cases, we have used the notation:
\begin{equation}
\mathcal{F}_N = \exp\{iNQa\} \quad \text{and} \quad Q \in \left[-\frac{\pi}{Na},\frac{\pi}{Na}\right)
\end{equation}
and we shall also be using the {\it length} of each species, defined as:
\begin{equation}
a_i = \frac{\hslash^2}{m e_i^2}
\end{equation}
One can now proceed to the generalization of the matrix form for general
number $N$ of species just by defining the
following $2\negmedspace\times \negmedspace2$ matrices:
\begin{equation}
    \mathbf{E}\,=\,\begin{pmatrix}
        e^{ika} & e^{-ika} \\
        -ike^{ika} & ike^{-ika}
        \end{pmatrix} \quad;\quad
    \mathbf{A_j}\,=\,\begin{pmatrix}
        -1 & -1 \\
        \big(ik-\frac{2}{a_j}\big) & -\big(ik+\frac{2}{a_j}\big)
        \end{pmatrix}.
\label{ec:matrix}
\end{equation}
\vskip 0.3cm
The matrices for $N=2$ and $N=3$ species given by \eqref{ec:det2} and \eqref{ec:det3} can now be
put in a more compact form with the help of
the $\mathbf{E}$ and $\mathbf{A_j}$ as:
\begin{equation}
           \begin{pmatrix} 
        \mathbf{E} & \mathbf{A_1} \\
        \mathbf{A_2}\mathcal{F}_2 & \mathbf{E}
        \end{pmatrix}_{4\times4} 
    \quad ; \quad
            \begin{pmatrix}
        \mathbf{E} & \mathbf{A_1} & 0_{2\times2} \\
        0_{2\times2} & \mathbf{E} & \mathbf{A_2} \\
        \mathbf{A_3}\mathcal{F}_3 & 0_{2\times2} & \mathbf{E}
        \end{pmatrix}_{6\times6}
\end{equation}

It is now relatively simple to guess that the general form of a matrix for
$N$ species must be written as:
\begin{equation}
    \begin{pmatrix}
        \mathbf{E} & \mathbf{A_1} & 0_{2\times2} & \hdotsfor{3} & 0_{2\times2} \\
        0_{2\times2} & \mathbf{E} & \mathbf{A_2} & 0_{2\times2} & \hdotsfor{2} & 0_{2\times2} \\
        \vdots & 0_{2\times2} & \mathbf{E} & \mathbf{A_3} &
        0_{2\times2} & \hdotsfor{1} & 0_{2\times2} \\
        \vdots & \vdots & 0_{2\times2} & \ddots & \ddots & \ddots & \vdots \\
        \vdots & \vdots & \vdots & \ddots & \ddots & \ddots & 0_{2\times2} \\
        0_{2\times2}& \vdots & \vdots & & \ddots & \mathbf{E} &
        \mathbf{A_{N-1}} \\ 
        \mathcal{F}\mathbf{A_N} & 0_{2\times2} & 0_{2\times2} &\hdotsfor{2} &
        0_{2\times2} & \mathbf{E}   
    \end{pmatrix}_{2N\times2N}
    \label{ec:detgen}
\end{equation}

So far nothing very exciting has happened except that one can write the
matrices in a compact, logic and generalizable
way. And in fact without further steps the progress would have not been
certainly remarkable. The real breakthrough arises
when one realizes that one has to deal with the {\bf determinants equated
to zero of these matrices} in order to learn
something about the {\bf band condition  of this one dimensional N-species
quantum periodic structure}. Let us define the
following function  $(\epsilon=ka)$ \footnote{Notice that for negative energies, $\epsilon$ takes pure imaginary values 
 that we represent in the figures in the negative part of the spectrum}:
\begin{equation}
    h_i (\epsilon) = \cos (\epsilon) + \left(\frac{a}{a_i}\right)\frac{\sin (\epsilon)}{\epsilon}
\label{ec:h}
\end{equation}
A quite simple computer algebra calculation shows that the determinant
equated to zero of \eqref{ec:det2},  which belongs
to the $N=2$ case, can be written in terms of these functions as
\begin{equation}
    \cos(2Qa)= 2h_1h_2-1
\end{equation}
And the determinant of the $N=3$ matrix given by \eqref{ec:det3} can also be calculated
to yield:
\begin{equation}
    \cos(3Qa)= 4h_1h_2h_3-(h_1+h_2+h_3)
\end{equation}
Below,  the cases $N=4, 5, 6$ and 7 are explicitely listed, using the
generalized matrix \eqref{ec:detgen} for each case and
calculating the determinant equated to zero with the help of the $h(\epsilon)$
functions \eqref{ec:h}. The result is:
\begin{gather}
    \cos(4Qa) \,= \,8\;h_1 h_2 h_3 h_4 -2\;\big( h_1 h_2+h_1 h_4+h_2 h_3+ h_3
    h_4 \big) + 1  
\end{gather}
\begin{multline}
    \cos(5Qa) \,=\, 16\;h_1 h_2 h_3 h_4 h_5 -
    4\;\big(h_1h_2h_3+h_1h_2h_5+h_1h_4h_5 +\\ 
    h_2h_3h_4 + h_3h_4h_5 \big) + \big( h_1+h_2+h_3+h_4+h_5 \big) 
\end{multline} 
\begin{multline}
    \cos(6Qa) \,=\, 32\;h_1h_2h_3h_4h_5h_6 -
    8\,\big(h_1h_2h_3h_4+h_1h_2h_3h_6 + h_1h_2h_5h_6\,+\\
    + h_1h_4h_5h_6+h_2h_3h_4h_5+h_3h_4h_5h_6\big)+2\;\big(h_1h_2+h_1h_4+h_1h_6\,+\\
    + h_2h_3+h_2h_5+h_3h_4+h_3h_6+h_4h_5+h_5h_6\big)-1 
\end{multline} 
\begin{multline}
    \cos(7Qa) \,=\,64\;h_1h_2h_3h_4h_5h_6h_7-16\;\big(h_1h_2h_3h_4h_5+h_1h_2h_3h_4h_7\,+\\
    + h_1h_2h_3h_6h_7 +
    h_1h_2h_5h_6h_7+h_1h_4h_5h_6h_7+h_2h_3h_4h_5h_6+h_3h_4h_5h_6h_7\big)\,+\\
    +
    4\;\big(h_1h_2h_3+h_1h_2h_5+h_1h_2h_7+h_1h_4h_5+h_1h_4h_7+h_1h_6h_7+h_2h_3h_4\,+\\
    +h_2h_3h_6+h_2h_5h_6+h_3h_4h_5 + h_3h_4h_7+h_3h_6h_7+h_4h_5h_6+h_5h_6h_7\big) \,-\\
    - \big(h_1+h_2+h_3+h_4+h_5+h_6+h_7\big).
\end{multline} 

We have been able to proof by induction that the general form of the $N$
species case can be given as:
\begin{equation}
    \cos(NQa)\,=\,\mathcal{B}(\epsilon;a_1,\ldots,a_N)
\label{ec:condition}
\end{equation}
\begin{itemize}
\item  $N$ \textbf{even}\\[2mm]\hspace*{-5mm}
    \fbox{\begin{Beqnarray}
    \mathcal{B}(\epsilon;a_1,\ldots,a_N) = 2^{N-1}\sum_{P}h_i...(N)...h_k\, -\, 2^{N-3}\sum_{P}
    h_i...(N-2)...h_k \,+\, \notag \\ \label{ec:par}
    + \,2^{N-5}\sum_{P} h_i...(N-4)...h_k \,-\,\ldots (-1)^{\frac{N}{2}-1}\; 2 \sum_{P}
    h_i...(2)...h_k \,+\,
    (-1)^{N\over2}
    \end{Beqnarray}}
\item $N$ \textbf{odd}\\[2mm]\hspace*{-5mm}
    \fbox{\begin{Beqnarray}
    \mathcal{B}(\epsilon;a_1,\ldots,a_N) = 2^{N-1}\sum_{P}h_i...(N)...h_k \,-\, 2^{N-3}\sum_{P}
    h_i...(N-2)...h_k \,+ \notag \\ \label{ec:impar} +\, 2^{N-5}\sum_{P}
    h_i...(N-4)...h_k \,-\,\ldots (-1)^{\frac{N-3}{2}}\; 2^2 \sum_{P} h_i...(3)...h_k \,+\\
    (-1)^{N-1\over2}(h_1+h_2+h_3+\ldots+h_N) \notag
    \end{Beqnarray}}
\end{itemize}
All what remains is to define the symbol $\sum_{P} h_i...(M)...h_k$ which
means the {\it sum of all possible products of M
different $h_i$'s with the following rule for each product: the indices
must follow an increasing order and to an odd index must always follow an
even index and reciprocally.}

\begin{figure}[t]
    \epsfig{file=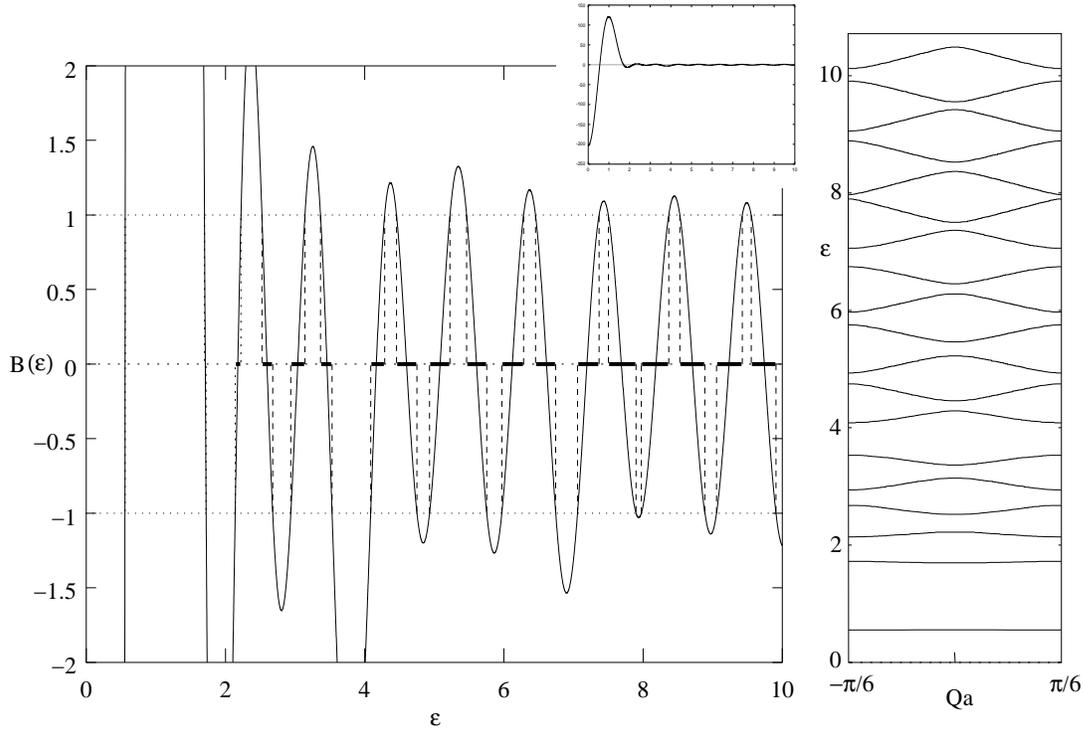,width=.9\textwidth}
\caption{Band condition (detail in inset) and band
structure (1BZ)
for a periodic chain with six species in the primitive cell 
$\big(\frac{a}{a_i}\big)$: 2, 2.5, 1, 1.5, -1, -2.}
\label{fig:six}
\end{figure}

\begin{figure}
    \epsfig{file=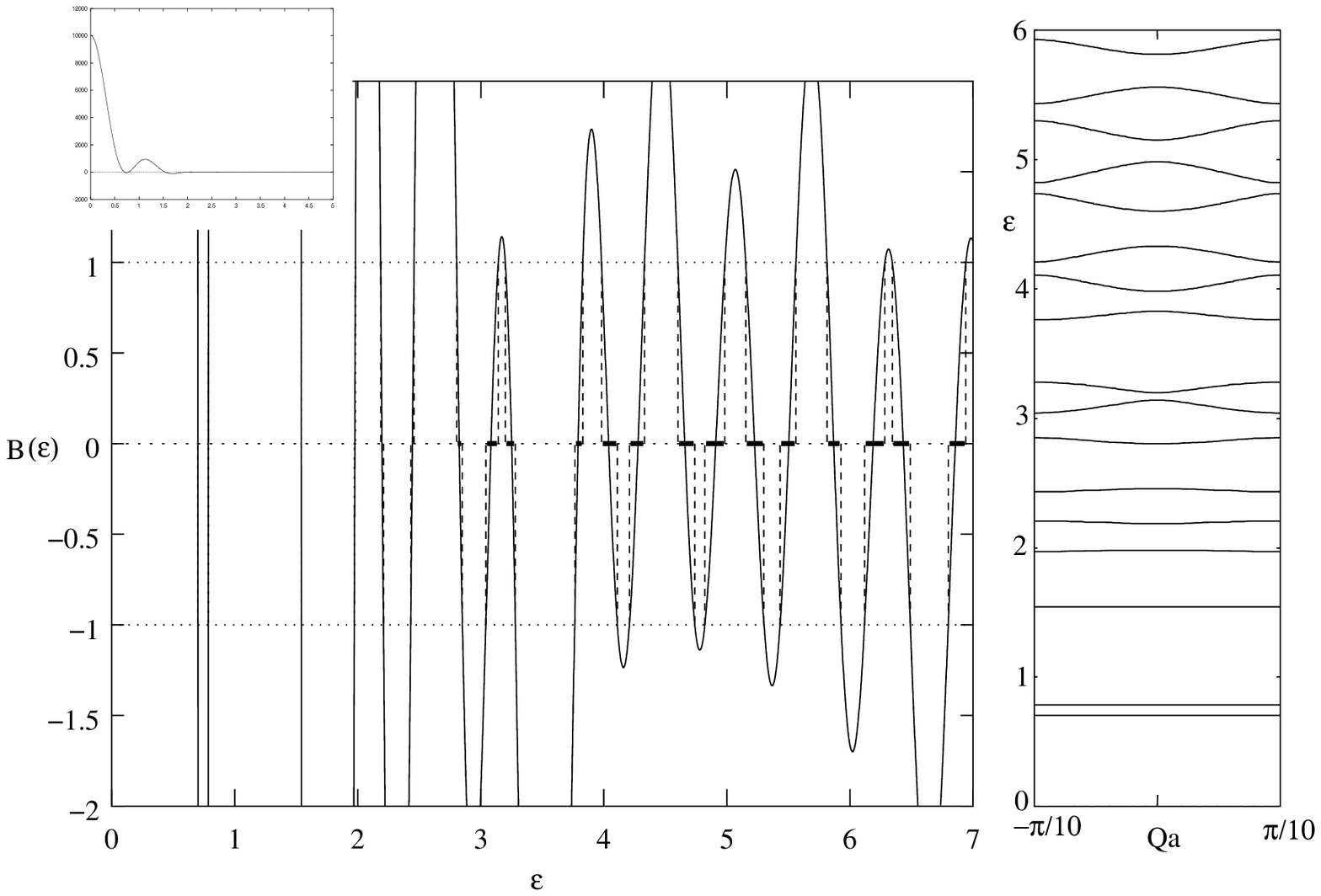,width=.9\textwidth}
\caption{Band condition (detail in inset) and band
structure (1BZ)
for a periodic chain with ten species in the primitive cell 
$\big(\frac{a}{a_i}\big)$: 1, 2, 3, -1, -2, -3, 0.5, -0.5, 2, 1.}
\label{fig:ten}
\end{figure}

The band structure provided by \eqref{ec:par} and \eqref{ec:impar} is not just {\bf exact} but
also extremely useful from the point of view
of computer algebra calculations. In fact we have carried out various
profiles for the curves provided for these
conditions until $N$=30 or more using just few seconds of a lap-top regular
computer. The reason for that lies mainly
in the systematic use of the form, products and combinations of the
$h(\epsilon)$-function defined by \eqref{ec:h}. As examples of
what has just been said we list in Fig.~\ref{fig:six}
and \ref{fig:ten} a series of band curves
for large number of species and various values
of the parameter $\left(\frac{a}{a_i}\right)$, defining the characteristic value of
the $h(\epsilon)$-function. 
\begin{figure}
\centering
    \subfigure[Two species $\big(\frac{a}{a_i}\big)$: -0.3,-0.9]{\epsfig{file=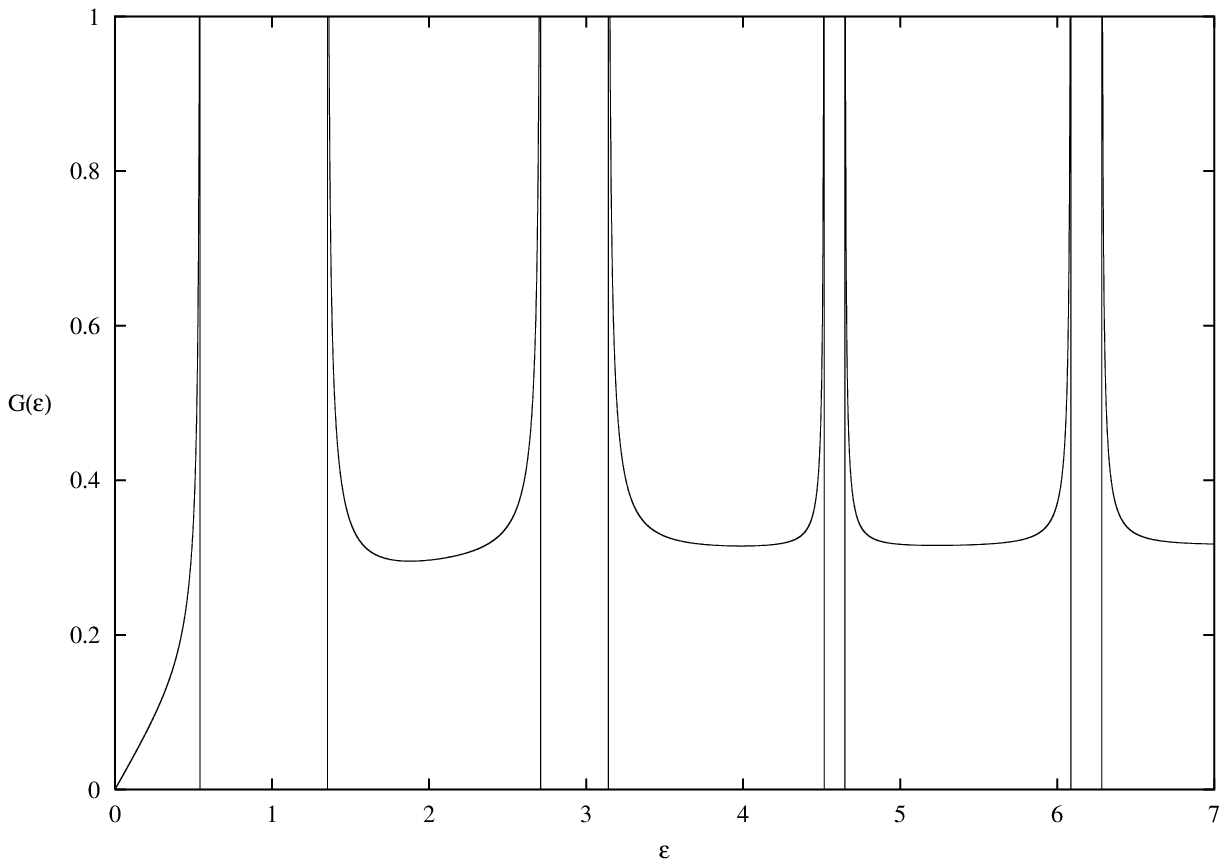,width=0.49\textwidth}}
    \subfigure[Four species $\big(\frac{a}{a_i}\big)$: -3, -5, -6, -4]{\epsfig{file=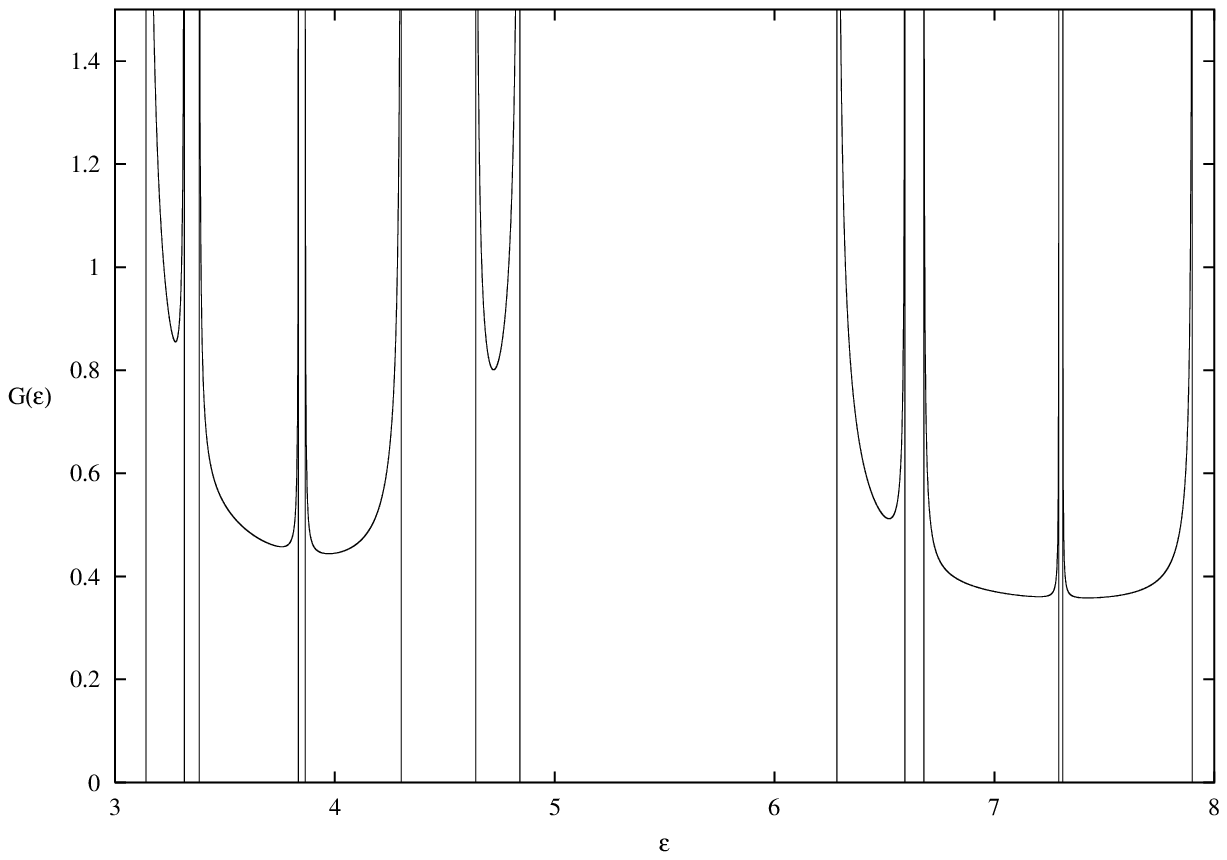,width=0.49\textwidth}}
\caption{Density of electronic states for different configurations of the
primitive cell.}
\label{fig:den}
\end{figure}
One can observe also the unpredictable set of allowed bands which appear at
low energies. This pattern increases its unpredictibility with the number
of species.

Once the band condition is known, one can write the distribution of
electronic states in a very simple form. In one dimension the density of
states per unit length of the chain for the n$th$ band comes from,
\begin{equation}
    g_n(\epsilon)\,=\,\frac{1}{2\pi}\sum_{Q}\bigg|\frac{d\epsilon(Q)}{dQ}\bigg|^{-1}
\end{equation}
where the sum is over all the first Brillouin zone (1BZ)  points $Q$
with the same energy $\epsilon$. Due to the parity of $\cos(NQa)$ the number of
points with the same value of $\epsilon$ in the 1BZ is always 2, and provided that 
overlapping of neighbouring bands is not possible in this system, we can write the density of
states as
\begin{equation}
    g(\epsilon)\,=\,\frac{1}{\pi}\bigg|\frac{dQ(\epsilon)}{d\epsilon}\bigg|
\end{equation}
inside the permitted bands. From \eqref{ec:condition} a trivial
calculation leads to,
\begin{equation}
    G(\epsilon)\equiv g(\epsilon)\cdot
a\,=\,\frac{\left[1-\mathcal{B}^2\right]^{-\frac{1}{2}}}{N\pi}\bigg|\frac{d\mathcal{B}(\epsilon)}{d\epsilon}\bigg|
\end{equation}
where $G(\epsilon)$ is the density of electronic states per
atom. Fig.~\ref{fig:den} shows some examples of the
characteristic form of the distribution of states for different
configurations of the primitive cell.

\section{Random Chains}

The structures one can observe in Nature hardly show a perfect periodicity.
 Even in the laboratory it is a difficult task to grow a crystal
free of impurities, vacancies or dislocations.
 We shall now treat the presence
of substitutional disorder in one dimensional delta-potential chains, that
is we consider a chain of equally spaced deltas in which the sequence of
different species does not obey a periodic pattern. This model has been
mainly studied regarding the vibrational spectrum \cite{LM},
paying less attention to its electronic density of states
\cite{AgaBor}. For the purpose of studying quantum wires the relevant
behaviour we want to analyze lies more in the latter than in the former
physical property. 

\subsection{Energy gaps}
The Saxon and Hutner conjecture \cite{SH} was proved 
by Luttinger for the case of  binary chains \cite{lutt}. We have been able
to extend it to the general case.
A detailed calculation following the line of Schmidt \cite{Schmidt} can be
found in Appendix \ref{ap:gaps}. The result can be easily summarized as follows:
{\bf the forbidden
bands that coincide in different one species delta chains with couplings
$e_1^2,\ldots,e_N^2$ are also forbidden levels in any infinite
chain made up of deltas of the $N$ types.} 
This conclusion can be applied to both disordered and periodic chains.
As can be seen in the calculations the result requires the interatomic
distances of the chains involved to be constant and the same for all of them.

\subsection{Eigenenergies for finite disorder chains}

Let us calculate the allowed energy levels of a
finite non-periodic chain of deltas with fixed end-points boundary
conditions.
The procedure to follow is the same as for the periodic case
 but imposing the vanishing of the
wave function at the end-points of the chain which we locate at one atomic
distance to the left of the first delta and to the right of the last one.
The connection of the wave function throughout the different sectors leads us
to a condition for the permitted energy levels in the form of the
determinant of a matrix equated to zero. Again these matrices can be
written in a generalizable way. Thus for $N$ deltas we have
\begin{equation}
    \begin{vmatrix}
        \mathbf{E} & \mathbf{A_1} & 0_{2\times2} & \hdotsfor{3} & 0_{2\times2} \\
        0_{2\times2} & \mathbf{E} & \mathbf{A_2} & 0_{2\times2} & \hdotsfor{2} & 0_{2\times2} \\
        \vdots & 0_{2\times2} & \mathbf{E} & \mathbf{A_3} &
        0_{2\times2} & \hdotsfor{1} & 0_{2\times2} \\
        \vdots & \vdots & 0_{2\times2} & \ddots & \ddots & \ddots & \vdots \\
        \vdots & \vdots & \vdots & \ddots & \ddots & \ddots & 0_{2\times2} \\
        0_{2\times2}& \vdots & \vdots & & \ddots & \mathbf{E} &
        \mathbf{A_N} \\ 
        \mathbf{U} & 0_{2\times2} & 0_{2\times2} &\hdotsfor{2} &
        0_{2\times2} & \mathbf{V}   
    \end{vmatrix}_{2(N+1)\times2(N+1)}\!=\,0.
\end{equation} 
whith $\mathbf{E}$ and $\mathbf{A_j}$ defined in \eqref{ec:matrix} and
\begin{equation}
    \mathbf{U}\,=\,\begin{pmatrix}
        1 & 1 \\ 0 & 0
        \end{pmatrix} 
    \qquad \mathbf{V}\,=\,\begin{pmatrix}
        0 & 0 \\ e^{ika} & e^{-ika}
        \end{pmatrix}.
\end{equation} 
We found the condition to be factorizable in terms of the functions
$h_j(\epsilon)$ in a similar manner to that of the periodic chain. The
eigenenergies of the system are the roots of:
\begin{equation}
    \sin(\epsilon)\cdot \mathcal{A}(\epsilon;a_1,\ldots,a_N)\,=\,0  
\label{ec:condicion}
\end{equation}
where
\begin{itemize}
\item  $N$ \textbf{even}\\[2mm]\hspace*{-5mm}
    \fbox{\begin{Beqnarray}
    \mathcal{A}(\epsilon;a_1,\ldots,a_N) = 2^N (h_1\cdot\ldots\cdot h_N)\, -\, 2^{N-2}\sideset{}{'}\sum_{P}
    h_i...(N-2)...h_k \,+\, \notag \\ \label{ec:despar}
    &\hspace*{-120mm}\displaystyle +\,2^{N-4}\sideset{}{'}\sum_{P}
h_i...(N-4)...h_k \,-\,\ldots (-1)^{\frac{N}{2}-1}\; 2^2 
\sideset{}{'}\sum_{P}
    h_i...(2)...h_k \,+\,
    (-1)^{\frac{N}{2}}
    \end{Beqnarray}}
\item $N$ \textbf{odd}\\[2mm]\hspace*{-5mm}
    \fbox{\begin{Beqnarray}
    \mathcal{A}(\epsilon;a_1,\ldots,a_N) = 2^{N-1} (h_1\cdot\ldots\cdot h_N) \,-\, 2^{N-3}\sideset{}{'}\sum_{P}
    h_i...(N-2)...h_k \,+ \notag \\ \label{ec:desimpar} +\, 2^{N-5}\sideset{}{'}\sum_{P}
    h_i...(N-4)...h_k \,-\,\ldots (-1)^{\frac{N-3}{2}}\; 2^2 \sideset{}{'}\sum_{P} h_i...(3)...h_k \,+\\
    (-1)^{\frac{N-1}{2}}(h_1+h_3+h_5+\ldots+h_{N-2}+h_N) \notag
    \end{Beqnarray}}
\end{itemize}
and here the symbol $\sideset{}{'}\sum_{P} h_i...(M)...h_k$ 
means the {\it sum of all possible products of M
different $h_i$'s with the following rule for each product: the first index has to be odd, 
the indices must follow an increasing order and to an odd index must always follow an
even index and reciprocally.}

From \eqref{ec:condicion} we see that  $\epsilon=n\pi \,;\,
n\in\mathbb{N}$, are always eigenvalues of any finite length disordered
chain wathever the species in it.

\subsection{Density of electronic states for an infinite chain}
The basic assumption concerning the electronic energies of a random chain
is that the distribution of levels converges to a limiting distribution as
the number of atoms goes to infinity, which is the same for almost all
atomic sequences (except for a fraction that goes to zero as
$N\rightarrow\infty$) as long as the concentrations of the different
species remain fixed. The existence of this property is needed by 
the thermodynamic limit: all  quantities charaterizing
macroscopically a random infinite chain with fixed species and
concentrations cannot depend on the order
of any finite-length piece of it.
The method used to obtain the density of states in random chains is
essentially due to James and Ginzbarg \cite{JG} and Schmidt \cite{Schmidt},
who derived a functional equation which supplies the limiting distribution
for the positive part of the spectrum.
We have reconstructed and extended the method to provide the density of
states for all energies ( Appendix \ref{ap:functional} ).
The functional equation is,
\begin{subequations}
\label{ec:w}
\begin{align}
    & \mathbf{W}(\varphi)\,=\,\sum_{i=1}^m
    p_i\left\{\mathbf{W}\left(\mathcal{T}^{-1}_i(\varphi)\right)-\mathbf{W}\left(\mathcal{T}^{-1}_i(0)\right)
    \right\}\\
    & \mathbf{W}(\varphi +r\pi)\,=\,\mathbf{W}(\varphi)+r \\
    & \mathbf{W}(\pi)\,=\,1 \\
    & \mathbf{W}(\varphi) \text{ is  monotonically increasing in } \varphi 
\end{align}
\end{subequations}
where $\varphi\in(0,\pi]$ and $m$ is the number of species composing the chain
, $p_i$ the $i$th species concentration and
$\mathcal{T}^{-1}_i(\varphi)$ are the functions,
\begin{equation}
    \mathcal{T}^{-1}_i(\varphi)\,=\,
        \arctan\left(2h_i(\epsilon)-\dfrac{1}{\tan(\varphi)}\right) 
\end{equation}
with  $h_i(\epsilon)$ defined in \eqref{ec:h}.
Solving \eqref{ec:w} for different values of the energy, the density of
electronic states per atom can be obtained from,
\begin{equation}
G(\epsilon)\,=\,\frac{\left| \mathcal{K}\left(\epsilon+\frac{\Delta
\epsilon}{2}\right)-\mathcal{K}\left(\epsilon-\frac{\Delta
\epsilon}{2}\right)\right|}{\Delta \epsilon}
\end{equation}
where
$\mathcal{K}=-\sum_{i=1}^mp_i\left[\mathbf{W}\left(\mathcal{T}^{-1}_i(0)\right)\right]$.

\begin{figure}
\vspace*{-10mm}
\subfigure[2 species. $\big(\frac{a}{a_i}\big)\{p_i\}:
-2\{0.5\},\,-0.25\{0.5\}$. The dotted line is the density of states for the
ordered diatomic chain]{\epsfig{file=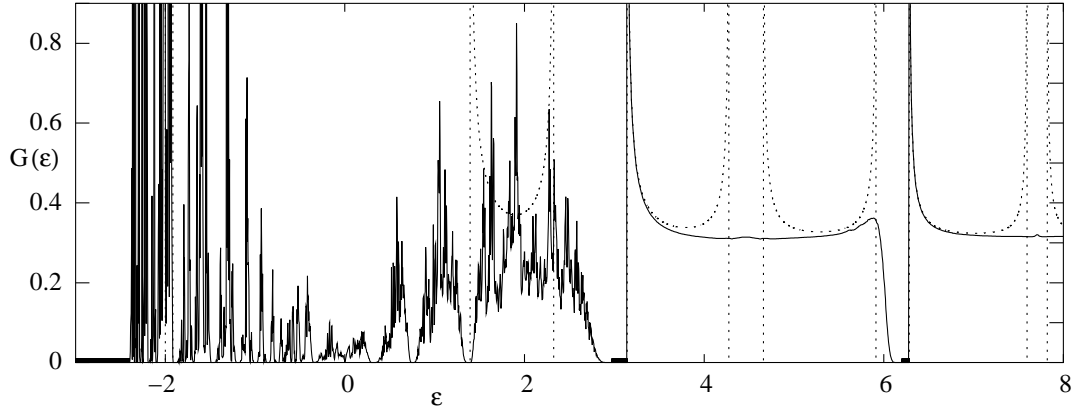,width=.9\textwidth}}\vspace*{-4mm}
\subfigure[3 species. $\big(\frac{a}{a_i}\big)\{p_i\}:
10\{0.3\},\,-1\{0.3\},\, -2\{0.4\}$] {\epsfig{file=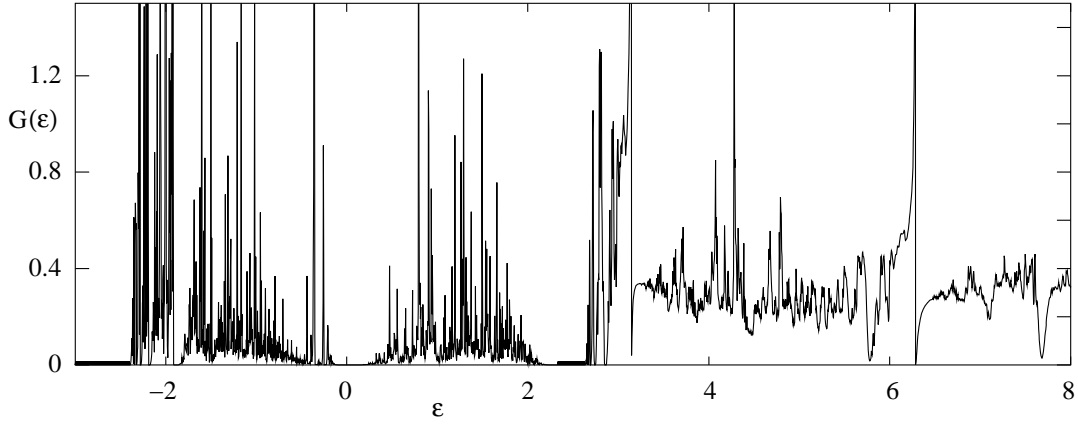,width=.9\textwidth}}\vspace*{-4mm}
\subfigure[8 species. $\big(\frac{a}{a_i}\big)\{p_i\}:
 -0.5\{0.1\},\, -1\{0.1\},\, -1.5\{0.1\},\,-2\{0.1\},\, -2.5\{0.1\},\,
-3\{0.1\},\, -3.5\{0.1\},\,4\{0.3\}$]{\epsfig{file=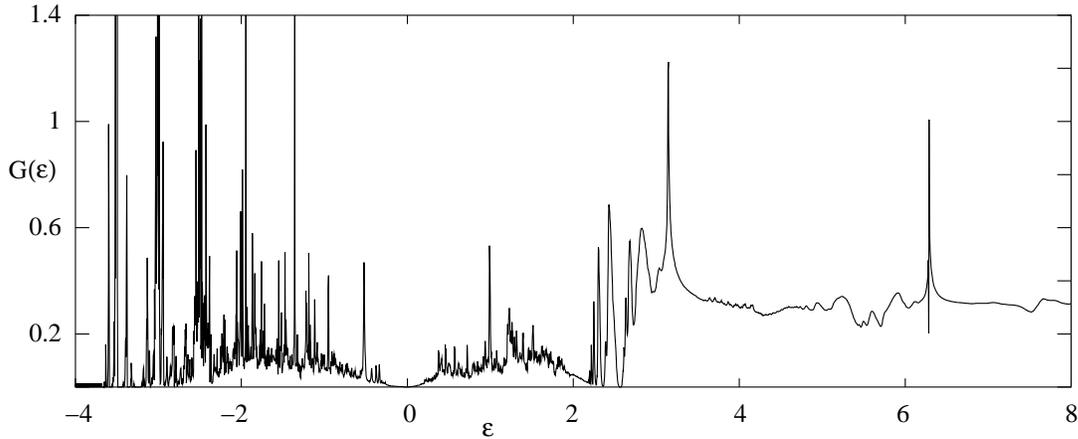,width=.9\textwidth}}
\caption{Distributions of electronic states for different configurations of
the random chain. The thick line over the abscisa axis marks the common
forbidden bands for the species composing the array. 5001 points have been taken for
$\varphi$ and $\Delta\epsilon=4\cdot 10^{-3}$.}
\label{fig:dens}
\end{figure}

We solve this equation numerically. The range of $\varphi$ is discretized
in $p$ equally spaced points and $\mathbf{W}$ is represented by its values
at those points which using a selfconsistent algorithm we have been able
to calculate up to an error of $10^{-15}$. Some examples of distributions
of random chains are shown in Fig.~\ref{fig:dens}.

The most interesting feature of the different spectra  is the localized
peak structure that appears for certain energy intervals. Those
irregularities had been reported several years ago
\cite{AgaBor}\cite{Taylor} and now the extremely accurate numerical
algorithms dramatically confirm. Although the general interpretation of
the behaviour of the distributions for these systems is non trivial one can
say as a
general rule (and following the suggestion of Agacy and Borland
\cite{AgaBor}) that the peaked regions are  remarkable in 
 ranges of the spectrum which are forbidden for some of the species involved
but not for all of them since in this latter case the range is also not
allowed for the random system. A clear view of this explanation can
be obtained by looking at a random chain of two species A and B. In a region
forbidden for the A-type chain the allowed energies  appearing in  the spectrum are
due to atomic clusters involving B atoms. As a result the A atoms
surrounding these groups isolate those energies causing 
the density of states to decrease when we move in a tiny region around each of the
levels and therefore giving rise to a fluctuating distribution.
\begin{figure}
\vspace*{-10mm}
\subfigure[$p_A=p_B=0.5$]{\epsfig{file=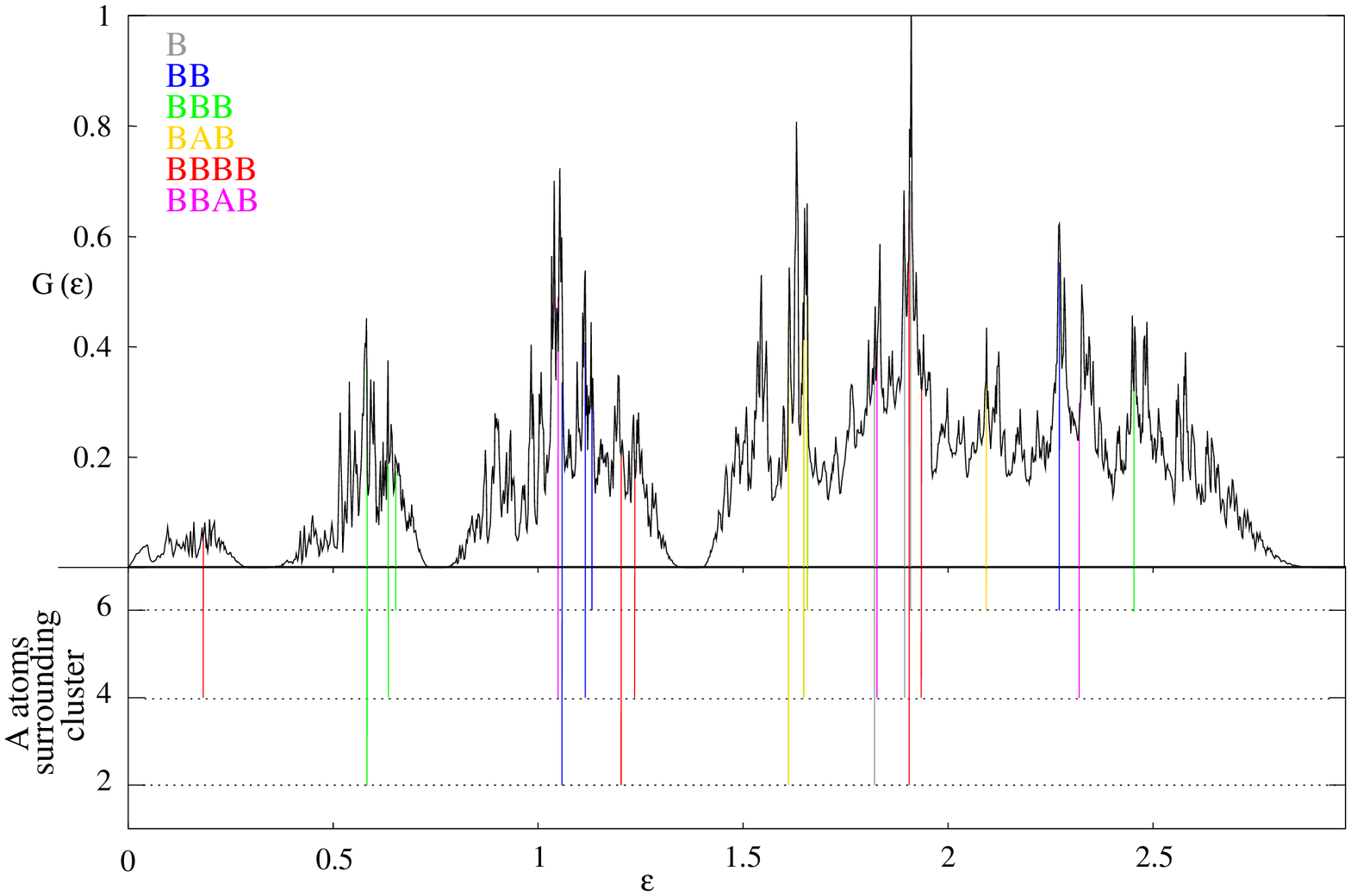,width=.9\textwidth}}
\subfigure[$p_A=0.9$, $p_B=0.1$]{\epsfig{file=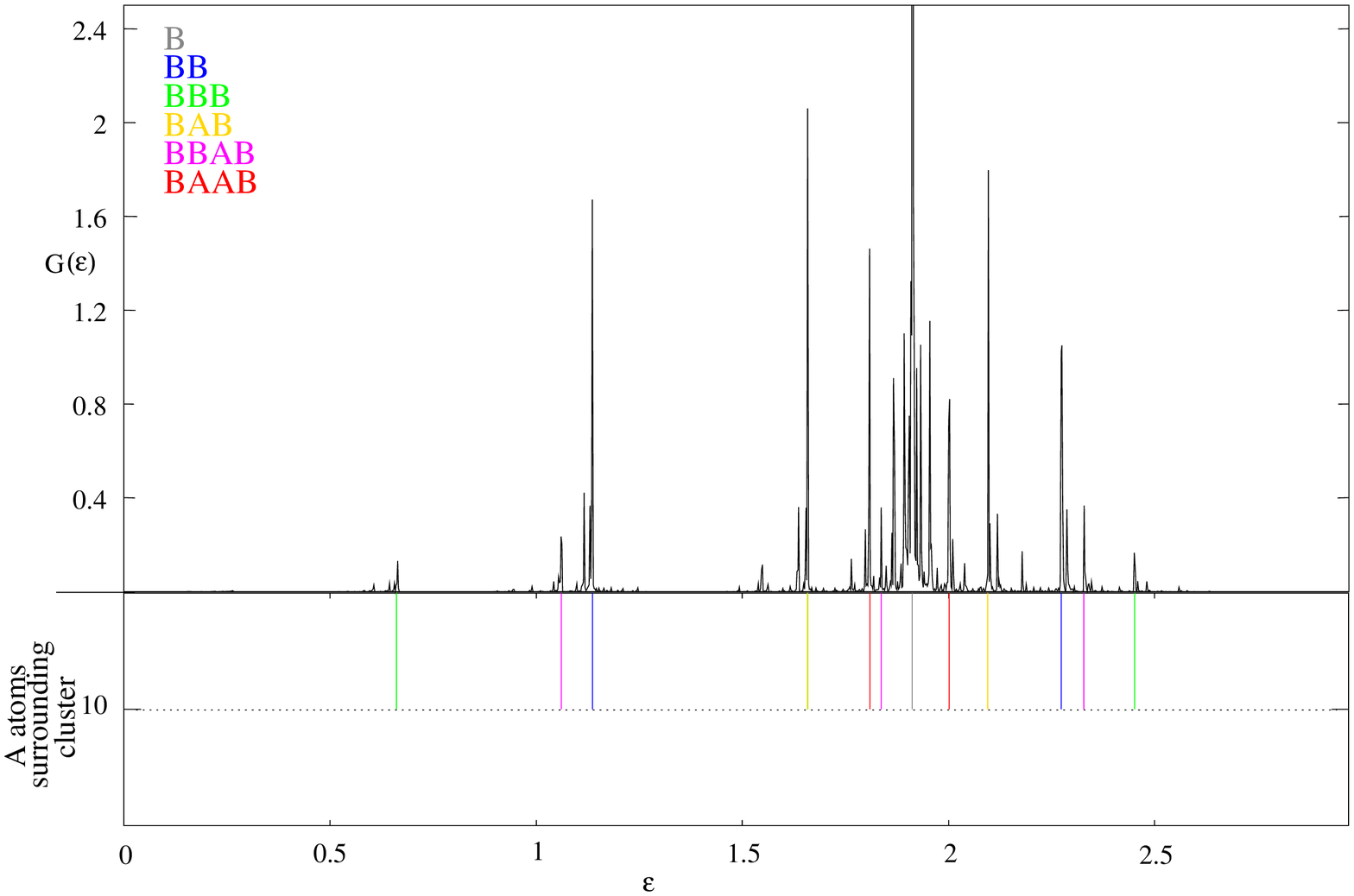,width=.9\textwidth}}
\caption{Density of states for a 2 species random chain with
$\big(\frac{a}{a_A}\big)=-2$ and $\big(\frac{a}{a_B}\big)=-0.25$ in a range
forbidden for the A-type chain. The color vertical lines indicate the position of
the eigenstates of different atomic B-groups surrounded by a certain number
of A atoms. 5001 points have been taken for
$\varphi$ and $\Delta\epsilon=2\cdot 10^{-3}$.}
\label{fig:clusters}
\end{figure}
Thus one could reproduce the energies where the density of states would be
more prominent, from the eigenvalues of certain atomic clusters in which
the B species have a substantial contribution as shown in Fig.~\ref{fig:clusters}. 
In the (a) example it is clear how the eigenstates clusterize around the
more peaked regions of the distribution. However one needs to consider a huge
number of clusters in order to reproduce all the maxima due to
the equal concentrations of the species. Decreasing the B concentration we
see in (b) how the sharp points can be quite easily predicted.

Another feature that can be seen  in the different representations in
Fig.~\ref{fig:dens}, is how the distribution approaches a smoother curve
that resembles the one corresponding to a periodic chain as the energy
grows: the greater the electron's energy  the less it feels the presence
of disorder.

\begin{figure}
\centering
\epsfig{file=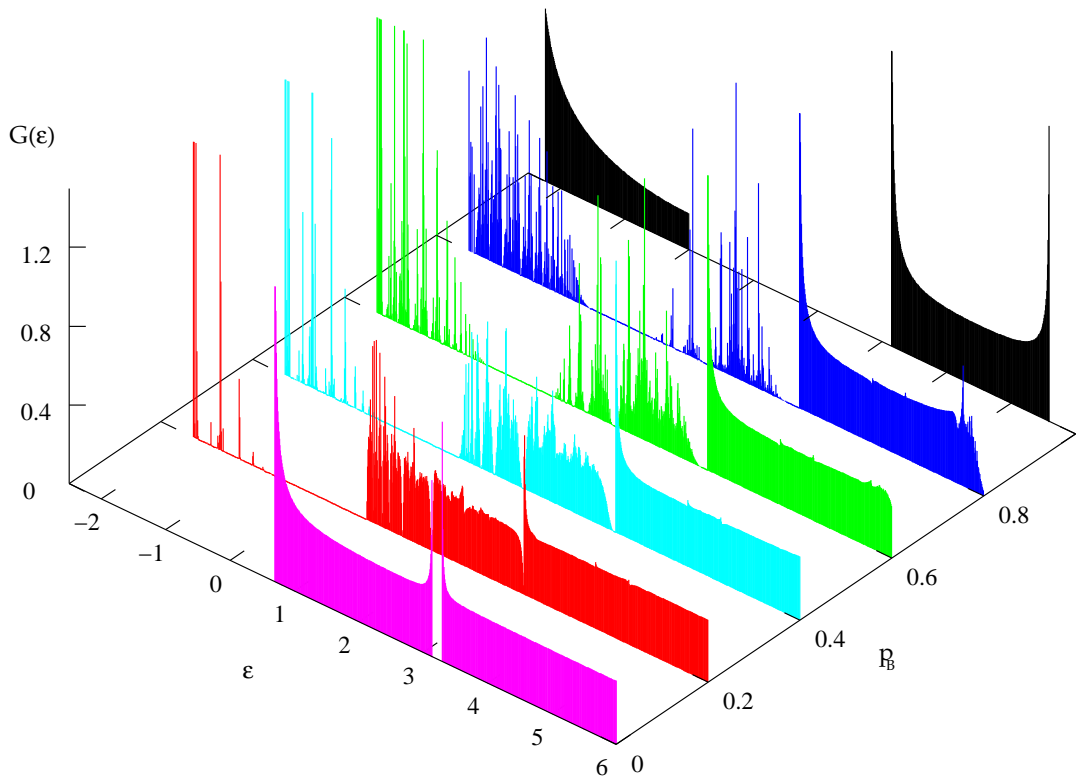,width=.9\textwidth}
\caption{Distribution of states for 2 species random chains with  
$\big(\frac{a}{a_A}\big)=0.25$ and \mbox{$\big(\frac{a}{a_B}\big)=-2$} in
different concentrations. 3001 points have been taken for $\varphi$ and $\Delta\epsilon=8\cdot 10^{-3}$.}
\label{fig:3D}
\end{figure}

In Fig.~\ref{fig:3D}, the evolution of the density of states can be
observed for a pure one species chain as we dope substitutionally with
atoms of a different kind.

\subsection{The degree of localization of the electronic states}

Since Anderson \cite{Anderson} the disorder in the structures has been
accepted as the main effect responsible of the spatial localization of the eigenstates of the systems.
Just  few years ago it was shown how if the disorder is correlated the
localization can be suppressed for certain energies
\cite{Dunlap}\cite{Sanchez}. In 1999 the first experimental evidence that
correlations inhibit localization of states in disordered low-dimensional
systems was reported \cite{Diez}.

Let us check the degree of localization for the electronic states in
 random chains with positive energy. What we  specifically calculate is the average
logarithmic decay per atom of the square of the envelope of the real wave
function which seems an adequate quantity for such a characterization
\cite{Borland}. 

\begin{figure}
\subfigure[2 species $\big(\frac{a}{a_1}\big)=-2$ and
$\big(\frac{a}{a_2}\big)=-0.25$ 
 with equal concentrations]{\epsfig{file=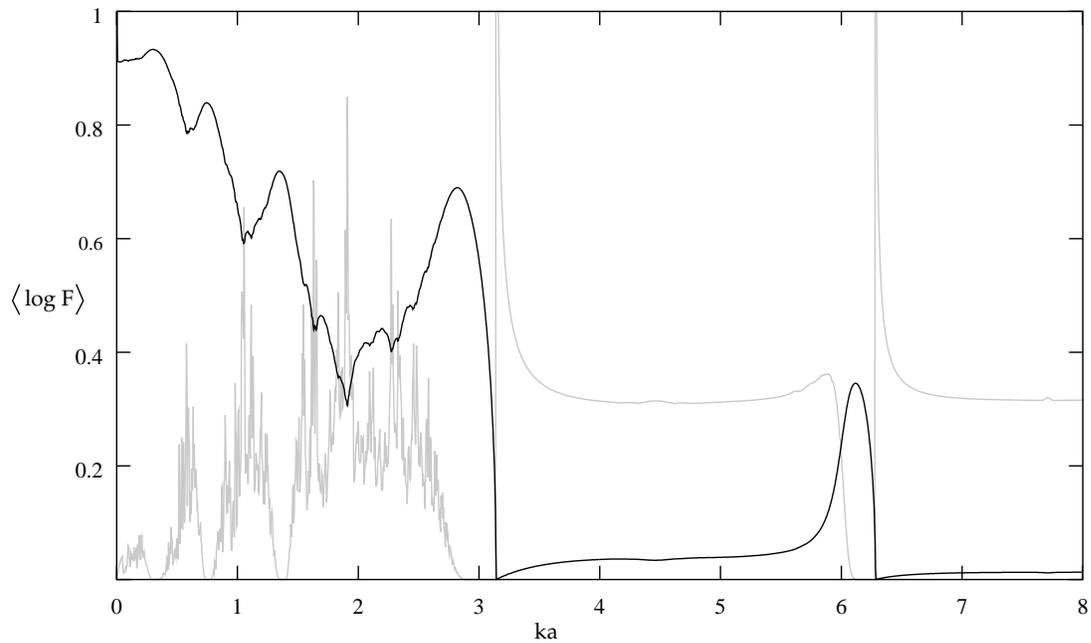,width=.9\textwidth}}
\subfigure[3 species $\big(\frac{a}{a_1}\big)=10$, $\big(\frac{a}{a_2}\big)=-1$ and
$\big(\frac{a}{a_3}\big)=-2$ with
$p_1=p_2=0.3$ and $p_3=0.4$]{\epsfig{file=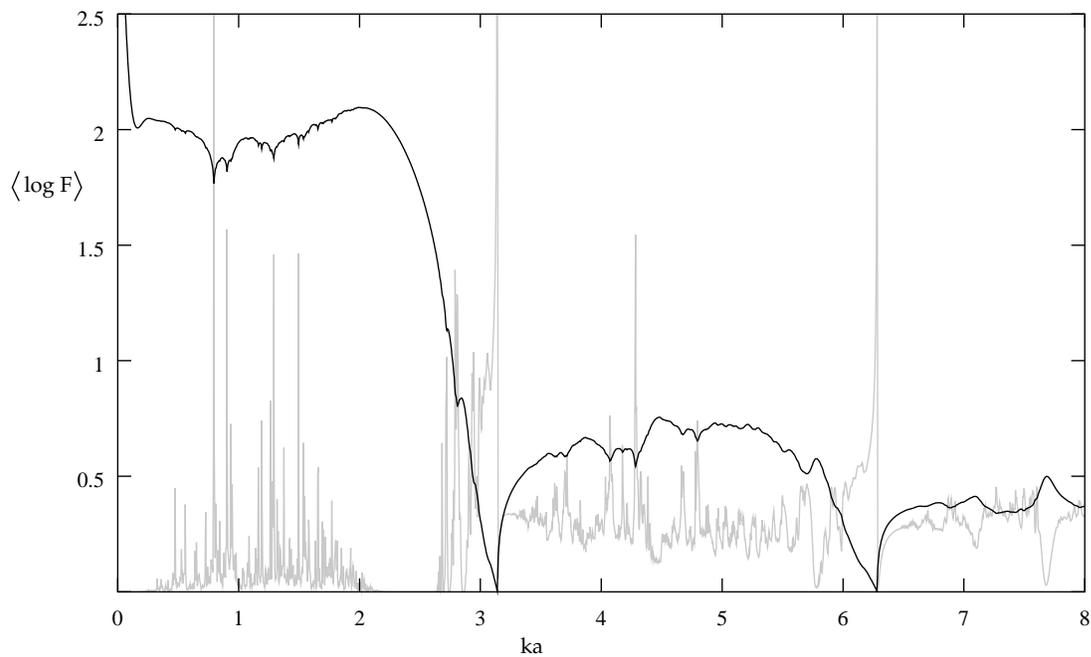,width=.9\textwidth}}
\caption{Degree of localization for the electronic states (black line)
and density of states (grey line) for a random
chain. $5001$ points have benn taken for $\varphi$ to solve the functional equation
and $\Delta(ka)=4\cdot 10^{-3}$.}
\label{fig:loca}
\end{figure}

In the $j$th sector we write:
\begin{equation}
    \Psi_j(x)\,=\,A_j\cos\big(kx-k(j-1)a+\phi_j\big)
\end{equation}
and we look for:
\begin{equation}
\lim_{N\rightarrow\infty}\frac{1}{N}\sum_{j=1}^N
\log\left(\frac{A_{j+1}^2}{A_j^2}\right).
\label{ec:pro}
\end{equation}
The connection equations of the wave function at the borders of the
different sectors imply:
\begin{equation}
\frac{A_{j+1}^2}{A_j^2}\,=\,1+
\left(\frac{2}{ka_j}\right)^2\cos^2(ka+\phi_j)-\left(\frac{2}{ka_j}\right)\sin(2ka+2\phi_j).
\label{ec:c2loca}
\end{equation}
Again, we will be using the functional equation to carry out the calculation
so using the phase $\varphi_j$ as in the Appendix
\ref{ap:functional}, it is just a matter of algebra to obtain:
\begin{multline}
F(\varphi_j,a_j)\,\equiv\,\frac{A_{j+1}^2}{A_j^2}\,=\,1+
\left(\frac{2}{ka_j}\right)^2\frac{\sin^2(ka)}{1-2\cos(ka)\tan(\varphi_j)+\tan^2(\varphi_j)}\,+\\
+\,\left(\frac{2}{ka_j}\right)\frac{2\sin(ka)\big(\cos(ka)-\tan(\varphi_j)\big)}{1-2\cos(ka)\tan(\varphi_j)+\tan^2(\varphi_j)}
\end{multline}
and the same arguments apearing on  the Appendix lead us to write 
\eqref{ec:pro} as
\begin{equation}
  \langle \log F \rangle\,=\,\sum_{i=1}^m p_i\int_0^{\pi}
\frac{d\mathbf{W}(\varphi)}{d\varphi}\log F(\varphi,a_i) d\varphi.
\end{equation} 
Integrating by parts and using the equations for $\mathbf{W}(\varphi)$, we
finally obtain: 
\begin{equation}
   \langle \log F \rangle\,=\,\sum_{i=1}^m p_i\log F(\pi,a_i)\,-\,\sum_{i=1}^m
p_i\int_0^{\pi}\mathbf{W}(\varphi)\frac{1}{F(\varphi,a_i)}\frac{dF(\varphi,a_i)}{d\varphi}
d \varphi.
\end{equation}
In the Fig.~\ref{fig:loca} some examples of the degree of localization as
a function of the energy are shown.
As can be seen the degree of localization has a tendency to decrease in
the peaks of the spectrum and increase in the troughs, and for $ka=n\pi\,;\,
 n\in\mathbb{N}$ the states are always extended for any chain. This
last result had been already obtained using the trasfer matrix technique
\cite{Azbel} for finite chains. So the system with completely
uncorrelated disorder can support extented states. In fact an infinite
number of isolated resonances is present altough mobility
edge for the electrons does not exist and 
therefore one cannot speak of an Anderson transition in a strict sense for
this particularly simple model. 

\section{Fractality}

\begin{figure}
    \epsfig{file=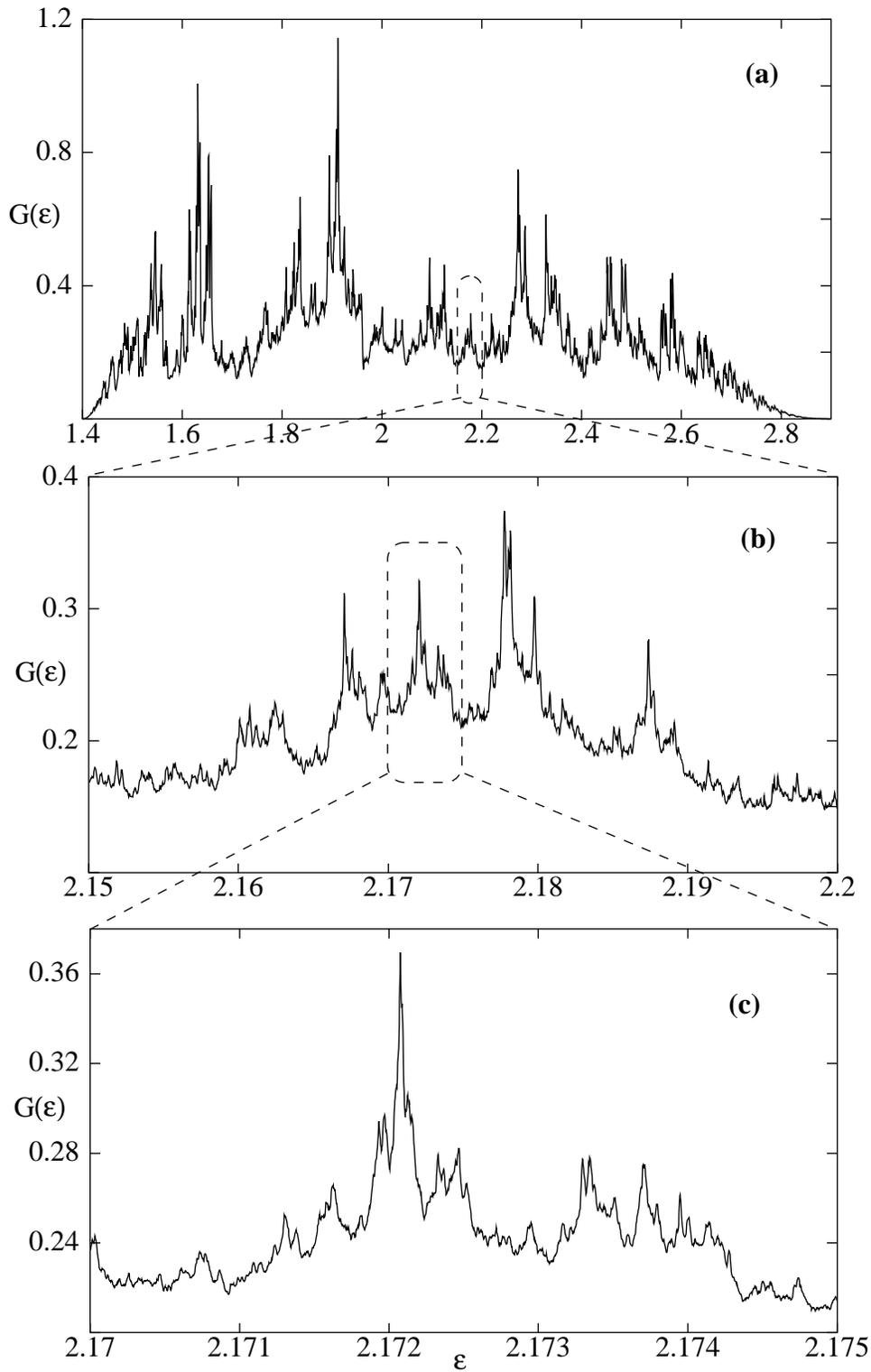, height=\textheight}
    \caption{Density of electronic states in different energy ranges for a
two species random chain $\big(\frac{a}{a_1}\big)=-2$ and
$\big(\frac{a}{a_2}\big)=-0.25$ with equal concentrations (fractal region).}
    \label{fig:fractal}
\end{figure}

\begin{figure}
    \epsfig{file=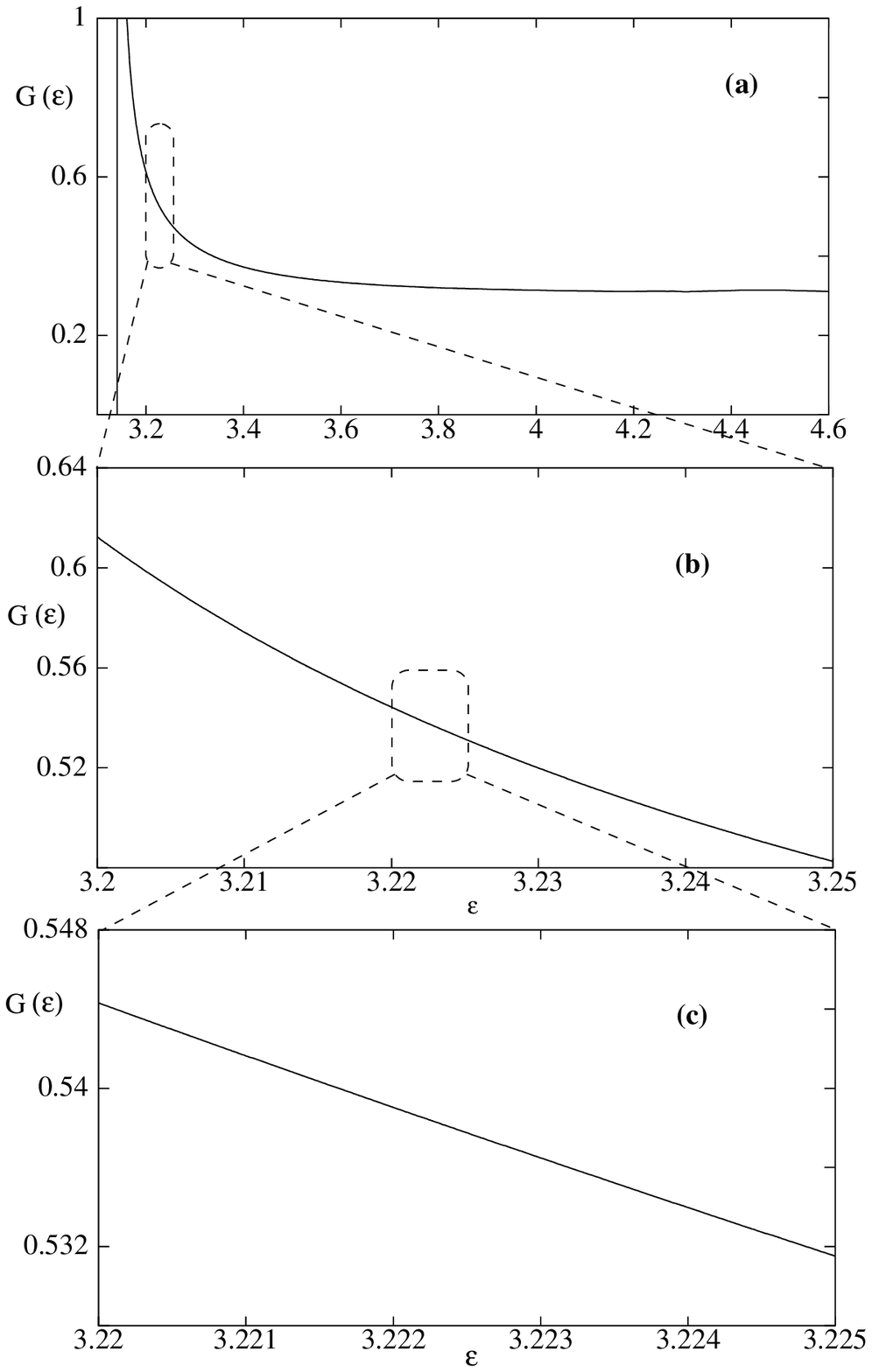, height=\textheight}
    \caption{Density of electronic states in different energy ranges for a
two species random chain $\big(\frac{a}{a_1}\big)=-2$ and
$\big(\frac{a}{a_2}\big)=-0.25$ with equal concentrations smooth region).}
    \label{fig:nofractal}
\end{figure}

Let us observe the peaked regions of the distributions. At first
sight the irregularities make the density of states appear hardly differentiable
within those intervals. Would this pattern still hold as we look deeper
into the distribution? Is the distribution non differentiable? In order to
answer these questions one has to improve the numerical calculation to be
able to see the real density of states in smaller energy ranges. In
Fig.~\ref{fig:fractal}
the spectrum of a certain random chain is shown for shorter and shorter
energy intervals. 
As can be seen as the energy domain is made smaller and thus increasing the accuracy of
the numerical algorithm the distribution reveals a finer structure: new
sharp points appear and the density does not evolve smoothly. We have been
extremely careful with the numerical algorithms and we are pretty sure that the
observed irregularities are not due to numerical errors. For the
representations in Fig.~\ref{fig:fractal} we have proceeded doubling  the
number of points taken for $\varphi$ in order to represent the functional
equation and checking the convergence of the density of states at each step
until the desired accuracy ( in all cases the average variation of
$G(\epsilon)$ in the last step relative to its domain was less than 0.75\%
). The final parameters were:
\begin{center}
\setlength{\extrarowheight}{3pt}
\begin{tabular}{c c c}
 & points for $\varphi$ & $\Delta \epsilon$ \\ \hline \hline   
Fig. \ref{fig:fractal} (a) & 5001 & $7.5 \cdot 10^{-4}$\\
Fig. \ref{fig:fractal} (b) & 35001 & $2.5 \cdot 10^{-5}$\\
Fig. \ref{fig:fractal} (c) & 150001 & $2.5 \cdot 10^{-6}$\\\hline\\[-2mm]
\end{tabular}
\end{center}
As a final check we repeated the procedure for the same random chain, with
the same paremeters but in an apparently smooth region of the density of
states obtaining the results shown in Fig.~\ref{fig:nofractal}. From this
we can see that the irregularities in certain intervals of the spectrum are not due to
errors of the numerical computation. 

\begin{figure}[hb]
    \centering
    \epsfig{file=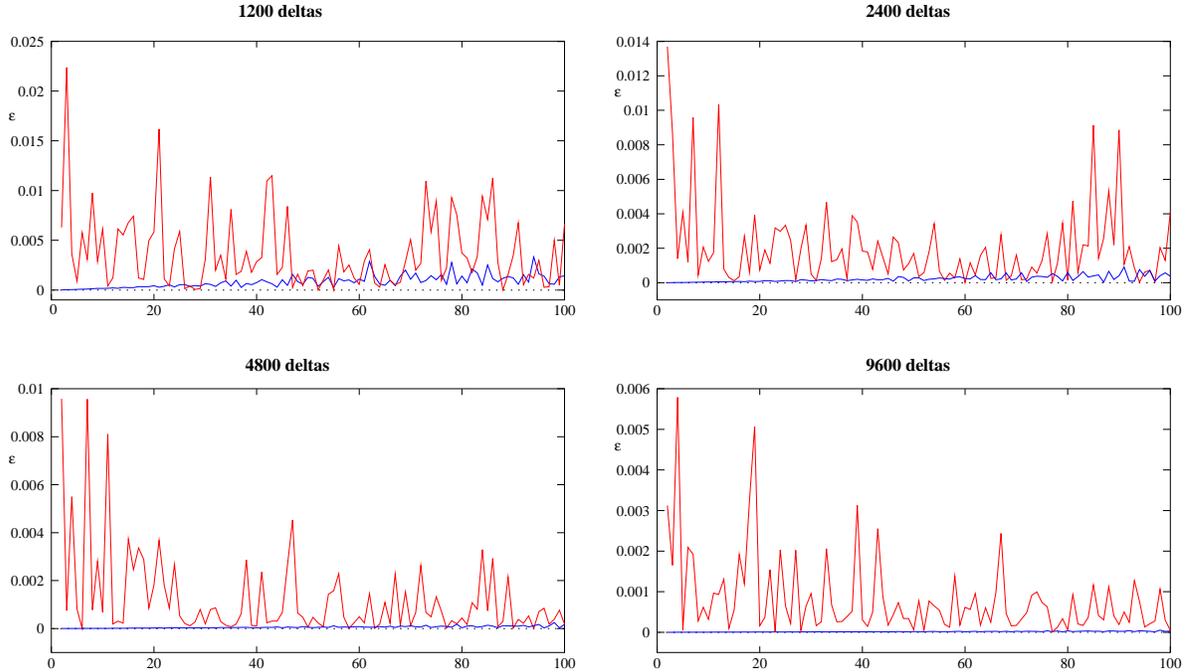,width=\textwidth}
    \caption{Energy spacings for the first 100 eigenvalues appearing inside
an irregular zone ( $\epsilon > 1.4$, red ) and inside a regular one (
$\epsilon > \pi$, blue ). The point $n$ on the abscisa axis represents the energy distance
between the $(n-1)$ and $n$ level. All the sequences are composed of the
species $\big(\frac{a}{a_1}\big)=-2$ and
$\big(\frac{a}{a_2}\big)=-0.25$ with equal concentrations. For a fixed length all the
states calculated correspond to the same random sequence. The states have
been calculated by finding the transmission of the wave function through the
chain with fixed end-points boundary conditions.}
    \label{fig:niveles}
\end{figure}

The numerical calculation of the density of states suggests the possibility
for the distributions to show a quasi-fractal behaviour in some energy
ranges: non differentiability for any point inside these regions and irregular
aspect whatever the scale. The differentiability of the distribution is a
consequence of a regular (almost homogeneous) distribution of the states
inside a small energy interval. A clear view of the differences between the
way in which  states appear inside an irregular region and a smooth region of the
density can be obtained representing the energy spaces for adjacent
levels. In Fig.~\ref{fig:niveles} these spacing distributions are shown for
the first 100 levels that appear inside an irregular zone and a smooth
one. Increasing the length of the random chain we see how the spacing
distribution for the levels in the smooth region becomes more and more
homogeneous as the first 100 levels are included in a smaller energy
interval while the spacings for the levels in the irregular zone 
does not show a defined tendency nor a homogeneus distribution. In fact
these last spacing distributions exhibit the same aspect whatever the 
scale would be.

The fractality of the density of states does not 
depend on any particular parameter of the random chain; its presence is a
consequence of the disorder and the dimensionality of the system and it is in
this sense a universal effect. This fractal behaviour might be related to
the fractal conductance fluctuations observed in gold nanowires \cite{Hegger} in the
mesoscopic regime suggesting a connection between disorder and coherent
transport.

\section{Concluding remarks}
    To summarize we would like to point out the main results we have
obtained in this paper. First, we emphasize that the 
band structure provided by \eqref{ec:condition}-\eqref{ec:impar} is not just {\bf exact} but
also extremely useful from the point of view
of computer algebra calculations. In fact we have carried out various
profiles for the curves provided for these
conditions until $N$=30 or more using just few seconds of a lap-top regular
computer. The reason for that lies mainly
in the systematic use of the form, products and combinations of the
$h(\epsilon)$-function defined by \eqref{ec:h}. The Luttinger theorem has
been generalized to an arbitrary number of different delta potentials and
the Saxon-Hutner Conjecture has also been definitively established. Similar
expressions have been obtained for the case of a disordered finite chain
and the extended functional equation method has been used to look
specifically to the density of electronic states for a random infinite
array. By obtaining the degree of localization of the electronic states in
the random chains, we
have a more complete picture of the role the levels play in the transport 
processes and the existence of extended states under uncorrelated disorder has
been confirmed. The results are not only in agreement with what was
expected in more sophisticated models but also go beyond them and we can
account for universal properties of the transport effects by detecting
fractality in the conductance. This work is far from being finished as we
want to ascertain ourselves whether this effect can be measurable in terms
of the fractal dimension and critical exponents. At the moment, however, we
can already offer a large bundle of properties that arise from a model
whose simplicity is not only a shortcoming but rather an advantange to
study complex properties in  a benchproof easily manageable.

\appendix
\section{Appendix: Gaps theorem}
    \label{ap:gaps}
\begin{figure}[h]
    \centering
    \epsfig{file=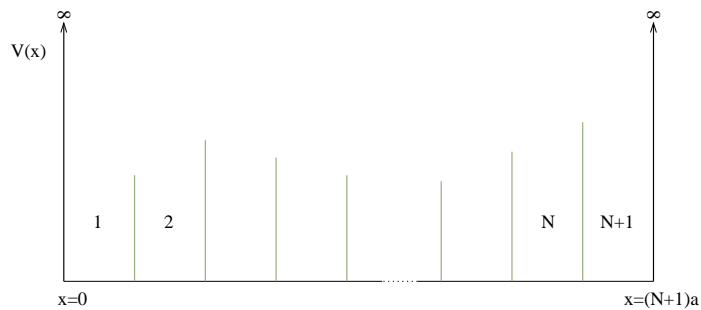,width=.6\textwidth}
    \caption{Deltas chain}
    \label{fig:box}
\end{figure}

Let us consider a finite chain of $N$ delta potentials of different species
with fixed end-points
boundary conditions ( Fig.~\ref{fig:box} ).
Inside the $j$th sector we write the wave function of the electron
\begin{equation}
    \Psi_j(x)\,=\,A_j e^{ikx_j}\,+\,B_j e^{-ikx_j},
\label{ec:psi}
\end{equation}
where $x_j=x-(j-1)a$.

Imposing the conditions at the border $x=ja$ one can obtain the
relationship between the amplitudes of adjacent sectors:
\begin{equation}
    \begin{pmatrix}
    A_{j+1} \\ B_{j+1} \end{pmatrix}\,=\,
    \underset{M}{\underbrace{\begin{pmatrix}
        \big(1-\frac{i}{ka_j}\big)e^{ika} & \frac{-i}{ka_j}e^{-ika}\\  
        \frac{i}{ka_j}e^{ika} & \big(1+\frac{i}{ka_j}\big)e^{-ika}
    \end{pmatrix}}}
    \begin{pmatrix}
    A_j \\ B_j \end{pmatrix}
    \label{ec:trans}
\end{equation}
and $M$ is the transmission matrix of the sectors.
The boundary conditions are:
\begin{subequations}
\label{ec:cont1}
\begin{gather}
        A_1\,+\,B_1\,=\,0 \\
    A_{N+1}e^{ika}\,+\,B_{N+1}e^{-ika}\,=\,0. 
\end{gather}
\end{subequations}
Let us define inside each sector
$\tau_j=-ie^{ika}\dfrac{A_j}{B_j}$. The Eqs.~\eqref{ec:cont1} become:
\begin{subequations}
\label{ec:contau}
\begin{gather}
    \tau_1\,=\,i e^{ika} \\
    \tau_{N+1}\,=\,i e^{-ika} ,
\end{gather}
\end{subequations}
and from \eqref{ec:trans} the transmission for $\tau_j$ is
\begin{equation}
    \tau_{j+1}\,=\,\frac{c_j\tau_j+1}{\tau_j+c_j^*}
    \label{ec:tauj+1}
\end{equation}
where $c_j=(i-ka_j)e^{ika}$. Let us use  $z_j=\dfrac{e^{ika}+i
\tau_j}{1+i\tau_j e^{ika}}$. The boundary conditions now read:
\begin{subequations}
\label{ec:contz}
\begin{gather}
   z_1\,=\,0 \\
    z_{N+1}\,=\,\infty .
\end{gather}
\end{subequations}
A straightforward calculation using \eqref{ec:tauj+1} lead us to:
\begin{equation}
    z_{j+1}\,=\,\frac{1}{2h_j(ka)-z_j}.
\label{ec:zj+1}
\end{equation}
with $h_j(ka)=\cos(ka)+\left(\frac{a}{a_j}\right)\dfrac{\sin(ka)}{ka}$.
Now let us consider the values of $ka$ satisfying $|h_j(ka)|>1$ for
every $a_j$ in the chain (i.e. the common forbidden bands for all the one
species chains). As $|z_1|<1$ it is easy to proof by induction using
\eqref{ec:zj+1} that
$|z_j|<1 \Rightarrow |z_{j+1}|<1$. So we have $|z_{N+1}|<1$  for arbitrary
$N$ and the final
boundary condition cannot be satisfied whatever the length of the
chain. Therefore those $ka$ values are not allowed in the mixed chain.
Notice that the order in which the different species appear is irrelevant.
The same conclusion holds for both periodic and disordered chains.

The result is also valid for complex values of $k$ (i.e. negative energies).

\section{Appendix: Functional equation}
\label{ap:functional}
The limiting distribution of electronic states for a random chain composed
of $m$ different species each one with concentration $p_i$, will be given by the
average of the spectra for all sequences which have the given
concentrations in the limit as the number of atoms goes to infinity.

Let us consider the same chain as in the Appendix \ref{ap:gaps} and the
quantities $z_j$. Since $z_1=0$ and the transmission \eqref{ec:zj+1} is real
every $z_j$ will also be real and we can define a phase in each sector
$\varphi_j$ so that $\tan(\varphi_j)=z_j$, which yields the result:
\begin{equation}
    \varphi_{j+1}\equiv \mathcal{T}_j(\varphi_j)\,=\,\arctan\left(\frac{1}{2h_j(\epsilon)-\tan(\varphi_j)}\right).
\end{equation}
We need the transmission of the phase to be an increasing continuous
function. Thus we define:
\begin{gather}
    \mathcal{T}_j(\varphi_j)\,=\,\begin{cases}
    \arctan\left(\frac{1}{2h_j(\epsilon)-\tan(\varphi_j)}\right) \quad   & \text{si }
\varphi_j\in\Big(-\frac{\pi}{2},\arctan\big(2h_j(\epsilon)\big)\Big]\\
    \ \arctan\left(\frac{1}{2h_j(\epsilon)-\tan(\varphi_j)}\right) +\pi \quad & \text{si
}\varphi_j\in\Big(\arctan\big(2h_j(\epsilon)\big),\frac{\pi}{2}\Big]\\
    \end{cases}
    \label{ec:apECasig1}\\
    \mathcal{T}_j(\varphi_j+n\pi)\,=\,\mathcal{T}_j(\varphi_j)+n\pi \qquad
\varphi_j \in \left(-\frac{\pi}{2},\frac{\pi}{2}\right].
\label{ec:apECasig2}
\end{gather}
 This can be easily written by using the inverse function as:
\begin{gather}
    \mathcal{T}^{-1}_j(\varphi_j)\,=\,\arctan\left(2h_j(\epsilon)-\frac{1}{\tan(\varphi_j)}\right) \label{ec:apECasig3}\\
    \mathcal{T}^{-1}_j(\varphi_j+n\pi)\,=\,\mathcal{T}^{-1}_j(\varphi_j)+n\pi \qquad
\varphi_j \in \left(0,\pi\right].
\label{ec:apECasig4}
\end{gather}
The boundary conditions are:
\begin{align}
    \varphi_1 &= 0 \label{ec:cc}\\
    \varphi_{N+1} &= \frac{\pi}{2}+n\pi \qquad n\in\mathbb{Z}.
\end{align}
Once the transmission of the phases can be written in a uniquely way, 
$\varphi_{N+1}(\epsilon)$ is a continuous function of $\epsilon$. Let us
assume the latter to be an increasing function. Then if
$\epsilon_1<\epsilon_2$ the quantity 
 $ \dfrac{\varphi_{N+1}(\epsilon_2)-\varphi_{N+1}(\epsilon_1)}{\pi}$
is the number of times the final boundary condition has been satisfied
from $\epsilon_1$ to $\epsilon_2$ with
an error smaller than 1 and therefore it also represents the number of eigenstates in the
interval $\epsilon_1<\epsilon< \epsilon_2$. So it is clear that we can
write the density of states per atom of the chain whatever the behaviour
of the function $\varphi_{N+1}(\epsilon)$ would be as:
\begin{equation}
    G(\epsilon)\,=\,\frac{1}{\pi}\frac{\left|\frac{\varphi_{N+1}(\epsilon+d\epsilon)}{N}-\frac{\varphi_{N+1}(\epsilon)}{N}\right|}{d\epsilon}
\end{equation}
As \eqref{ec:cc} holds for all $\epsilon$, we can write $\frac{\varphi_{N+1}(\epsilon)}{N}$:
\begin{equation}
    \frac{\varphi_{N+1}}{N}\,=\,\frac{1}{N}\sum_{j=1}^N\big[\varphi_{j+1}-\varphi_j\big]\,=\,\sum_{j=1}^N
    \frac{\mathcal{T}_j(\varphi_j)-\varphi_j}{N}
\end{equation}
for each value of $\epsilon$. Thus $\frac{\varphi_{N+1}(\epsilon)}{N}$ is
the average over all atoms of the advanced phase which we denote 
$\langle\Delta\varphi\rangle (\epsilon)$. 
 As $\mathcal{T}_j(\varphi_j)-\varphi_j$ is a periodic function with period
$\pi$ the average $\langle\Delta\varphi\rangle$ can be calculated using
the distribution functions of the phases $\varphi_j\modu{\pi}$. 
\begin{equation}
  \langle\Delta\varphi\rangle\,=\,\frac{1}{N}\sum_{j=1}^N\int_0^{\pi}\frac{dW_j(\varphi)}{d\varphi}\left\{
      \mathcal{T}_j(\varphi)-\varphi\right\} d\varphi
\label{ec:apECinte}
\end{equation}
where $W_j(\varphi)$ with $\varphi\in\left(0,\pi\right]$ is the
probability that $\varphi_j\modu{\pi}$ 
 lies in the interval
$\left(0,\varphi\right]$. Also we impose:
\begin{equation}
    W_j(\varphi +r\pi)\,=\,W_j(\varphi)+r \qquad
\varphi\in\left(0,\pi\right].
\label{ec:apECdis}
\end{equation}
The probability distribution for $\varphi_j$ depends on the distribution
for $\varphi_{j-1}$: $\varphi_j\modu{\pi}$  lies in
$\left(0,\varphi\right]$ if and only if $\varphi_{j-1}\modu{\pi}$ lies in
$\left(\mathcal{T}^{-1}_j(0),\mathcal{T}^{-1}_j(\varphi)\right]\modu{\pi}$.
There exist integers $s$ and $t$ such that
\begin{gather}
    \mathcal{T}^{-1}_j(0)(\nnnegthickspace\mod\pi)\equiv \mathbf{T}^{-1}_j(0)\,=\,\mathcal{T}^{-1}_j(0) -s\pi \\
    \mathcal{T}^{-1}_j(\varphi)(\nnnegthickspace\mod\pi)\equiv \mathbf{T}^{-1}_j(\varphi)\,=\,\mathcal{T}^{-1}_j(\varphi) -t\pi
\end{gather}
and $0<\mathcal{T}^{-1}_j(\varphi)-\mathcal{T}^{-1}_j(0)\leq
\pi$ from \eqref{ec:apECasig3} thus $t$ must be equal to $s$ or $s+1$.
\begin{description}
\item[\quad 1.] \boxed{$t=s$}\\
In this case (Fig.~\ref{fig:tiguals}):\\ $W_j(\varphi)=W_{j-1}\left(\mathbf{T}^{-1}_j(\varphi)\right)-
W_{j-1}\left(\mathbf{T}^{-1}_j(0)\right)$.
\begin{figure}[h]
    \centering
    \epsfig{file=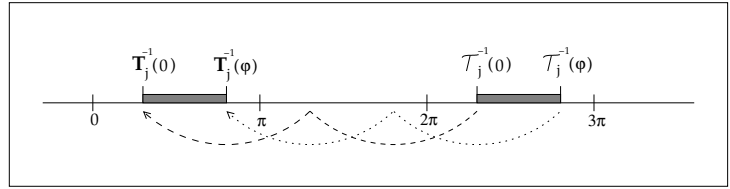,width=.6\textwidth}
    \caption{Example $t=s=2$}
    \label{fig:tiguals}
\end{figure}
\item[\quad 2.] \boxed{$t=s+1$}\\
In this case (Fig.~\ref{fig:tiguals1}):\\$W_j(\varphi)=1-W_{j-1}\left(\mathbf{T}^{-1}_j(0)\right)+
W_{j-1}\left(\mathbf{T}^{-1}_j(\varphi)\right)$. 
\begin{figure}[h]
    \centering
    \epsfig{file=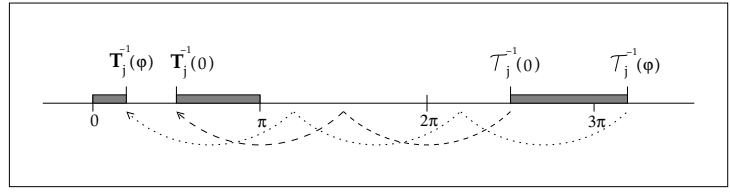,width=.6\textwidth}
    \caption{Example $s=2$ and $t=3$}
    \label{fig:tiguals1}
\end{figure}
\end{description}
Using \eqref{ec:apECdis} we can unify both cases as 
$W_j(\varphi)\,=\,W_{j-1}\left(\mathcal{T}^{-1}_j(\varphi)\right)-W_{j-1}\left(\mathcal{T}^{-1}_j(0)\right)$.
Thus the distribution functions are the solutions of the equations
\begin{subequations}
\label{ec:apECW}
\begin{align}
    & W_j(\varphi)\,=\,W_{j-1}\left(\mathcal{T}^{-1}_j(\varphi)\right)-W_{j-1}\left(\mathcal{T}^{-1}_j(0)\right)\\
    & W_j(\varphi +r\pi)\,=\,W_j(\varphi)+r \\
    & W_j(\pi)\,=\,1 \\
    & W_j(\varphi) \text{ is an increasing function of } \varphi 
\end{align}
\end{subequations}
with $\varphi\in (0,\pi]$.

The random chain is composed of $m$ different species with lengths
$a_1,\ldots,a_m$ in concentrations $p_1,\ldots,p_m$ ($\sum_{i=0}^m p_i=1$). 
We have now to take into account all possible sequences which have the
given concentrations. Thus we only know the probability of finding 
certain species in the position $x=ja$ and therefore we must take the
average. The Eq. \eqref{ec:apECinte} becomes:
\begin{equation}
    \langle\Delta\varphi\rangle\,=\,\frac{1}{N}\sum_{j=1}^N\int_0^{\pi}\frac{dW_j(\varphi)}{d\varphi}
    \sum_{i=1}^m p_i\left\{\mathcal{T}_i(\varphi)-\varphi\right\} d\varphi \,,
\label{ec:apECpro}
\end{equation}
and the equations for the distribution functions become:
\begin{subequations}
\label{ec:apECWW}
\begin{align}
    & W_j(\varphi)\,=\,\sum_{i=1}^m
    p_i\left\{W_{j-1}\left(\mathcal{T}^{-1}_i(\varphi)\right)-W_{j-1}\left(\mathcal{T}^{-1}_i(0)\right)
    \right\}\\
    & W_j(\varphi +r\pi)\,=\,W_j(\varphi)+r \\
    & W_j(\pi)\,=\,1 \\
    & W_j(\varphi) \text{ is an increasing function of } \varphi 
\end{align}
\end{subequations}
with $\varphi\in(0,\pi]$. Notice that now $W_j(\varphi)$ is the probability
that $\varphi_j\modu{\pi}$ lies in the interval $(0,\varphi]$ in some of
the possible sequences. Let us write \eqref{ec:apECpro} as
\begin{equation}
    \langle\Delta\varphi\rangle\,=\,\sum_{i=1}^m p_i\int_0^{\pi}
\frac{dW^N(\varphi)}{d\varphi}\left\{\mathcal{T}_i(\varphi)-\varphi\right\} d\varphi
\end{equation}
where $\displaystyle W^N(\varphi)\,=\,\frac{1}{N}\sum_{j=1}^N
W_j(\varphi)$, which is the average distribution function of the chain.
Let us take the limit $N\rightarrow \infty$ and denote $\displaystyle \mathbf{W}(\varphi)=\lim_{N\rightarrow\infty}
W^N(\varphi)$. It is not hard to see that this function will satisfy:
\begin{subequations}
\label{ec:apECw}
\begin{align}
    & \mathbf{W}(\varphi)\,=\,\sum_{i=1}^m
    p_i\left\{\mathbf{W}\left(\mathcal{T}^{-1}_i(\varphi)\right)-\mathbf{W}\left(\mathcal{T}^{-1}_i(0)\right)
    \right\}\label{ec:apECwa}\\
    & \mathbf{W}(\varphi +r\pi)\,=\,\mathbf{W}(\varphi)+r \label{ec:apECwb}\\
    & \mathbf{W}(\pi)\,=\,1 \\
    & \mathbf{W}(\varphi) \text{ is an increasing function of } \varphi 
\end{align}
\end{subequations}
with $\varphi\in(0,\pi]$. And,
\begin{equation}
    \langle\Delta\varphi\rangle\,=\,\sum_{i=1}^m p_i\int_0^{\pi}
\frac{d\mathbf{W}(\varphi)}{d\varphi}\left\{\mathcal{T}_i(\varphi)-\varphi\right\}
d\varphi .
\label{ec:apECincre}
\end{equation}
Schmidt has proved \cite{Schmidt} that the solution of Eqs.~\eqref{ec:apECw} 
for each energy is unique and continuous.
To carry out the integration in \eqref{ec:apECincre} we use the existence
of a value $\varphi_0=\frac{\pi}{2}$ such that $\mathcal{T}_i(\varphi_0)$ verifies $\mathcal{T}_i(\varphi_0)\equiv\varphi_1=\pi$ for
any species $a_i$.
\begin{equation*}
    \langle\Delta\varphi\rangle\,=\,\sum_{i=1}^m p_i\int_0^{\pi}
\frac{d\mathbf{W}(\varphi)}{d\varphi}\left\{\mathcal{T}_i(\varphi)-\varphi\right\}
d\varphi\,=\,\sum_{i=1}^m p_i\int_{\varphi_0}^{\varphi_0+\pi}
\frac{d\mathbf{W}(\varphi)}{d\varphi}\left\{\mathcal{T}_i(\varphi)-\varphi\right\}
d\varphi\,,
\end{equation*}
integrating by parts
\begin{align*}
   \langle\Delta\varphi\rangle = &\sum_{i=1}^m p_i \mathbf{W}(\varphi)\left\{
    \mathcal{T}_i(\varphi)-\varphi\right\}\bigg|_{\varphi_0}^{\varphi_0+\pi}\,-\,
    \sum_{i=1}^m
p_i\int_{\varphi_0}^{\varphi_0+\pi}\mathbf{W}(\varphi)\frac{d\mathcal{T}_i(\varphi)}{d\varphi}
    d\varphi \,+\\ & +\,\sum_{i=1}^m p_i \int_{\varphi_0}^{\varphi_0+\pi}\mathbf{W}(\varphi)d\varphi,
\end{align*}
and using \eqref{ec:apECwa}
\begin{multline*}
   \sum_{i=1}^m
p_i\int_{\varphi_0}^{\varphi_0+\pi}\mathbf{W}(\varphi)\frac{d\mathcal{T}_i(\varphi)}{d\varphi}
    d\varphi\,=\,\sum_{i=1}^m p_i
\int_{\varphi_1}^{\varphi_1+\pi}\mathbf{W}\left(\mathcal{T}_i^{-1}(\theta)\right)d\theta\,=\\
=\,\int_{\varphi_1}^{\varphi_1+\pi}\mathbf{W}(\theta)d\theta\,+\,\pi
\sum_{i=1}^m p_i \mathbf{W}\left(\mathcal{T}_i^{-1}(0)\right).
\end{multline*}
Going back to the expression for $\langle\Delta\varphi\rangle$ we obtain:
\begin{equation*}
    \langle\Delta\varphi\rangle\,=\,(\varphi_1-\varphi_0)-\pi\sum_{i=1}^m p_i
\mathbf{W}\left(\mathcal{T}_i^{-1}(0)\right)+\int_{\varphi_0}^{\varphi_0+\pi}\mathbf{W}(\varphi)d\varphi-
\int_{\varphi_1}^{\varphi_1+\pi}\mathbf{W}(\theta)d\theta.
\end{equation*}
From \eqref{ec:apECwb} is easy to see that
\begin{equation*}
   \int_{\varphi_0}^{\varphi_0+\pi}\mathbf{W}(\varphi)d\varphi-
\int_{\varphi_1}^{\varphi_1+\pi}\mathbf{W}(\theta)d\theta\,=\,\varphi_0-\varphi_1
\end{equation*} 
and finally the following
relationship holds for a certain value $\epsilon$ of the energy:
\begin{equation}
    \frac{\langle\Delta\varphi\rangle}{\pi}\equiv \mathcal{K}\,=\,-\sum_{i=1}^m p_i
\mathbf{W}\left(\mathcal{T}_i^{-1}(0)\right).
\end{equation}

Solving the Eqs.~\eqref{ec:apECw} for different values of $\epsilon$     
one can calculate the density of states of the random chain from
\begin{equation}
    G(\epsilon)\,=\,\frac{\left|\mathcal{K}
\left(\epsilon+\frac{\Delta \epsilon}{2}\right)-\mathcal{K}\left(\epsilon-\frac{\Delta \epsilon}{2}\right)\right|}{\Delta \epsilon}.
\end{equation}


\begin{thebibliography}{}%
\bibitem{KP} Kronig R. de L. and Penney W. G. {\it Proc. R. Soc. A} {\bf 130}, 499-513 (1931)%
\bibitem{LM} Lieb E. H. and Mattis D. C. {\it Mathematical Physics in one
Dimension}.  Academic Press. New York. 1966. 
We strongly suggest this book to the interested reader. It contains a large
amount of seminal reprinted
articles as well as huge number of crucial references.
\bibitem{AgaBor} Agacy R. L. and Borland R. E.  {\it Proc. Phys. Soc.} {\bf
84}, 1017-1026 (1964)
\bibitem{Schmidt} Schmidt H. {\it Phys. Rev.} {\bf 105}, 425-441 (1957)
\bibitem{JG} James H. and Ginzbarg A. {\it J. Phys. Chem.} {\bf 57}
,840-848 (1953)
\bibitem{Taylor} Gubernatis J. E. and Taylor P. L. {\it J. Phys. C} {\bf
4}, L95-L96 (1971)
\bibitem{lutt} Luttinger J. M. {\it Philips Res. Rep.} {\bf 6}, 303-310 (1951)
\bibitem{Anderson} Anderson P. W. {\it Phys. Rev.} {\bf 109}, 1492-1505 (1958)
\bibitem{Dunlap} Dunlap D. H., Wu H-L. and Phillips P. W. {\it
Phys. Rev. Lett.} {\bf 65}, 88-91 (1990)
\bibitem{Sanchez} S\'anchez A., Maci\'a E. and Dom\'{\i}nguez-Adame F. {\it
Phys. Rev B} {\bf 49}, 147-157 (1994)
\bibitem{Diez} Bellani V. {\it et al.} {\it Phys. Rev. Lett} {\bf 82}, 2159-2162 (1999)
\bibitem{Borland} Borland R. E. {\it Proc. Roy. Soc. A} {\bf 274}, 529-545 (1963) 
\bibitem{Azbel} Azbel M. Ya. and Soven P. {\it Phys. Rev. Lett} {\bf 49},
751-754 (1982)
\bibitem{altman} Altmann S. L. {\it Band Theory of Solids}  Oxford University
Press. 1991
\bibitem{SZML}	Szmulowicz F. {\it Eur. J. Phys.}  {\bf 18}, 392-397 (1997)
\bibitem{CDEK}	Dekker C. {\it Phys. Today}.   {\bf 52}, 22-29 (1999)
\bibitem{SH}	Saxon D. S. and Hutner R. A. {\it Philips Res. Rep.}  {\b
4}, 81-122 (1949)
\bibitem{Hegger} Helmut Hegger {\it et al}. {\it
Phys. Rev. Lett.} {\bf 77}, 3885-3888 (1996)
\bibitem{FLUG} Flugge S.  {\it Practical Quantum Mechanics}. Vol. 1, 64-68.
Springer Verlag. 1971
\end{thebibliography}
\end{document}